\documentclass[12pt]{article}

\usepackage{amsmath}
\usepackage{amssymb}
\usepackage{amscd}
\usepackage{pstricks}
\usepackage{pst-node}
\usepackage{hyperref}

\newcommand{\beq}{\begin{equation}}
\newcommand{\eeq}{\end{equation}}
\newcommand{\beqa}{\begin{eqnarray}}
\newcommand{\eeqa}{\end{eqnarray}}

\newcommand{\lam}{\lambda}

\newcommand{\ep}{\epsilon_1}
\newcommand{\es}{\epsilon_2}

\numberwithin{equation}{section}

\def\tr{\mathrm{tr}}
\def\SO{\mathrm{SO}}
\def\SU{\mathrm{SU}}
\def\SL{\mathrm{SL}}
\def\GL{\mathrm{GL}}

\def\U{\mathrm{U}}
\def\CP{\mathbb{CP}}
\def\cB{\mathcal{B}}
\def\cF{\mathcal{F}}
\def\cH{\mathcal{H}}
\def\cI{\mathcal{I}}
\def\cJ{\mathcal{J}}
\def\cM{\mathcal{M}}
\def\cN{\mathcal{N}}
\def\cP{\mathcal{P}}

\def\cV{\mathcal{V}}
\def\bC{\mathbb{C}}
\def\bR{\mathbb{R}}
\def\bZ{\mathbb{Z}}
\def\bL{\mathbb{L}}
\def\diag{\mathrm{diag}}
\def\ket#1{|#1\rangle}
\def\bra#1{\langle#1|}
\def\vev#1{\langle#1\rangle}
\let\hat\widehat

\begin{document}

\begin{titlepage}
\begin{flushright}
\normalsize

\medskip

May, 2011
\end{flushright}
\vfil

\bigskip

\begin{center}
\LARGE Instanton counting with  a surface operator\\
and the chain-saw quiver
\end{center}

\vfil
\medskip

\begin{center}
\def\thefootnote{\fnsymbol{footnote}}

Hiroaki Kanno$^\dagger$ and Yuji Tachikawa$^\ddagger$ \footnote[1]{on leave from IPMU, the University of Tokyo} 

\bigskip\bigskip

$^\dagger$ Graduate School of Mathematics and KMI\footnote[4]
{Kobayashi-Maskawa Institute for the Origin of Particles and the Universe}, \\
Nagoya University, Nagoya, 464-8602, Japan

\bigskip

$^\ddagger$ School of Natural Sciences, Institute for Advanced Study, \\
Princeton, New Jersey 08504, USA
\end{center}

\vfil
\bigskip

\begin{center}
{\bfseries abstract}
\end{center}

\bigskip

We describe the moduli space of  $\SU(N)$ instantons in the presence of a general surface operator of type $N=n_1+\cdots+n_M$ in terms of the representations of the so-called chain-saw quiver, which allows us to write down the instanton partition function as a summation over the fixed point contributions labeled by Young diagrams. We find that the instanton partition function depends on the ordering of $n_I$ which fixes a choice of the parabolic structure. This is in accord with the fact that the Verma module of the W-algebra also depends on the ordering of $n_I$. By explicit calculations, we check that the partition function agrees with the norm of a coherent state in the corresponding Verma module.

\vfill

\end{titlepage}

\setcounter{tocdepth}{2}
\tableofcontents

\section{Introduction}
The instanton partition function of four-dimensional $\cN=2$ gauge theory, as defined in \cite{Moore:1997dj,Nekrasov:2002qd}, is long known to have a close relationship with two-dimensional conformal field theory \cite{Losev:2003py,Nekrasov:2003rj}. 
Recently it was realized that the instanton partition function for the gauge group $G=\SU(N)$ is governed by the W-algebra symmetry $W_N$ \cite{Alday:2009aq,Wyllard:2009hg,Mironov:2009by}.
In particular, the partition function of the pure $\SU(N)$ gauge theory is given by the norm of a coherent state in the Verma module \cite{Gaiotto:2009ma,Marshakov:2009gn,Taki:2009zd}. 
The relation between instantons and W-algebras was further explored in e.g.~\cite{Bonelli:2009zp,Kanno:2009ga,Drukker:2010jp,Passerini:2010pr,Kanno:2010kj,Drukker:2010vg,Hollands:2010xa}.

The appearance of the W-symmetry, more natural for a two-dimensional theory, is better viewed from a higher-dimensional perspective. 
Consider the six-dimensional $\cN=(2,0)$ theory of type $A_{N-1}$ on $\bR_t \times S^1 \times \bR^4$, with $\bR_t$ as the time direction.
First compactify along $S^1$ to obtain five-dimensional super $\SU(N)$ Yang-Mills on $\bR_t \times \bR^4$.
The BPS sector of its dynamics can be reduced to the supersymmetric quantum mechanics on the instanton moduli space.
After Nekrasov's deformation, the Hilbert space of the BPS sector $\cH_\text{BPS}$ is the equivariant cohomology of the moduli space of instantons. The partition function is the norm of a state in $\cH_\text{BPS}$.
By exchanging the order of compactification it should also be possible to regard $\cH_\text{BPS}$ as the Hilbert space of $S^1$ compactification of the two-dimensional theory obtained by the compactification of the six-dimensional theory on $\bR^4$ with Nekrasov's deformation.  Then the $W_N$-algebra referred to above should be the symmetry of this two-dimensional theory.

The correspondence can be further generalized by adding  surface operators on the four-dimensional side.\footnote{Another generalization is to consider $\SO$ or $E$-type 6d theory, see e.g.~\cite{Hollands:2010xa}.}
A class of surface operators in $\SU(N)$ gauge theory is characterized by a partition $N=n_1+n_2+\cdots+n_M$, which we denote as $[n_I]$.
For the partition $[1,1,\ldots,1]$, it was found that the instanton partition function in the presence of the surface operator is governed by the affine symmetry $\hat{\SU}(N)$ \cite{Braverman:2004vv,Braverman:2004cr,Alday:2010vg,Kozcaz:2010yp}.
For the surface operator of a more general type,
it was conjectured \cite{Braverman:2010ef} that the two-dimensional symmetry is the W-algebra $W(\hat\SU(N),[n_I])$ obtained by the quantum Drinfeld-Sokolov reduction \cite{Bershadsky:1989mf,Feigin:1990pn,deBoer:1993iz} for the embedding $\rho_{[n_I]}:\SU(2)\to\SU(N)$ corresponding to the partition $[n_I]$,
and that the instanton partition function of the pure gauge theory in the presence of the surface operator is equal to the norm of a coherent state in a Verma module. 
A few checks of this conjecture were performed in \cite{Wyllard:2010rp,Wyllard:2010vi,Tachikawa:2011dz}.

The aim of this paper is to provide further checks of this proposal. We first describe the moduli space of instantons in the presence of a general surface operator in terms of the representations of a quiver, which mathematicians call the \emph{chain-saw quiver} \cite{FFNR,FR}. 
The crucial point is the equivalence between the instanton moduli space with a surface operator and the instanton moduli space on an orbifold $\bC\times (\bC/\bZ_M)$. 
Then the quiver describing the instanton moduli space is obtained by performing the $\bZ_M$ orbifold operation to the standard ADHM quiver. 
With the quiver description at hand, it is straightforward to enumerate and calculate the contribution from the fixed points on the moduli space under the action of the spacetime rotation and the gauge rotation, confirming the formula proposed in \cite{Wyllard:2010vi}.
We will see that the quiver and the fixed point formula depend on the ordering of $n_I$. 
Therefore, in this paper, we do \emph{not} assume $n_I\ge n_{I+1}$. When the ordering is ignored, we denote the data by $[n_I]$, called a \emph{partition} of $N$. 
When the ordering is meaningful, we denote the data by $(n_I)$, called a \emph{composition} of $N$, following the terminology in the combinatorics. 

We then compare the structure of the fixed points with the Verma module of the W-algebra. 
In order to define the Verma module, the W-algebra generators need to be split into creation and annihilation operators. 
Although there is essentially a unique choice of such splitting for $W(\hat\SU(N),[1,1,\ldots,1])=\hat \SU(N)$ and for $W(\hat\SU(N),[N])=W_N$, 
we will find that for general $W(\hat\SU(N),[n_I])$ the splitting depends on the ordering of  $n_i$, or equivalently on the composition $(n_I)$.  The coherent state condition needs to be imposed accordingly.  
Once this dependence on the composition $(n_I)$ of both the quiver and the Verma module is understood, we will find that the instanton partition function nicely agrees with the norm of the coherent state.

The rest of the paper is organized as follows. In Sec.~\ref{2}, we start by recalling the data defining the surface operator.
The equivalence of the instanton moduli space with a surface operator and the moduli space of orbifolded instantons is then explained. This equivalence is then used to construct a quiver describing the instanton moduli space with a surface operator from the standard ADHM quiver. Finally we study the structure of the fixed points using the resulting quiver, and write down the instanton partition function as a summation over fixed point contributions. 
In Sec.~\ref{3}, we first describe the structure of general W-algebras, and then study the structure of their Verma modules.
After comparing the graded dimension of the Verma module with the generating functions of the number of the  fixed points on the instanton moduli space, we determine the conditions we put on the coherent state. We report on the explicit checks performed using a computer.
We conclude with a few remarks in Sec.~\ref{4}. In the Appendix the structure of the W-algebra we use is determined explicitly.

\section{Instanton moduli space with a surface operator}\label{2}
\subsection{Defining data of a surface operator}\label{2.1}
The surface operator as a half-BPS object in $\cN=4$ supersymmetric 
Yang-Mills theory was introduced by Gukov and Witten \cite{Gukov:2006jk}.
We can employ almost the same definition of 
the surface operator also in ${\cal N}=2$ 
theories.
In this section we review the basic
properties of the surface operator following \cite{Gukov:2006jk}\footnote{See also \cite{Gukov:2007ck, Gaiotto:2009fs,Tan:2009he,Tan:2009qq,Tan:2010dk,Bruzzo:2010fk} 
for the surface operators in $\mathcal{N}=2$ theories.}.

The surface operator is defined by prescribing a singular behavior 
of the gauge field. 
Let us consider $\SU(N)$ gauge field $A_\mu$ on $\mathbb{R}^4 \simeq \mathbb{C}^2$ 
with complex  coordinates $(z, w)$
and put a surface operator at $w =0$, filling the $z$-plane. 
Let $(r,\theta)$ be the polar coordinates of the transverse  $w$-plane,  $w=re^{i\theta}$.
In the presence of the surface operator
the gauge field diverges as
\beq
A_\mu dx^\mu \sim \mathrm{diag}(\alpha_1, \alpha_2, \cdots, \alpha_N)~i d\theta, 
\label{diverge}
\eeq
near the surface operator. 
By a gauge transformation one can assume $2\pi > \alpha_i \geq \alpha_{i+1} \geq 0$. 
The data $\vec{\alpha} = (\alpha_1, \alpha_2, \cdots, \alpha_N)$ 
give an element of the Lie algebra of the Cartan subgroup $\U(1)^{N-1}$ of the gauge group 
$G=\SU(N)$.
The commutant of $\vec{\alpha}$ defines a subgroup $\mathbb{L} \subset G$
which is called the Levi subgroup. 
This means that the gauge group $G$ is broken to $\mathbb{L}$ on the surface. 

Suppose $\vec\alpha$ has the structure \begin{equation}
\vec\alpha=(
\underbrace{\alpha_{(1)},\ldots,\alpha_{(1)}}_{\text{$n_1$ times}},
\underbrace{\alpha_{(2)},\ldots,\alpha_{(2)}}_{\text{$n_2$ times}},
\ldots,
\underbrace{\alpha_{(M)},\ldots,\alpha_{(M)}}_{\text{$n_M$ times}} ),\label{ordered}
\end{equation}
where $\alpha_{(I)}>\alpha_{(I+1)}$. 
Here $(n_I)$ is a composition of $N$.
Then the Levi subgroup is \begin{equation}
\bL=S[\U(n_1)\times \U(n_2)\times \cdots \times \U(n_M)].
\end{equation}
Note that the Levi subgroup of $\SU(N)$ only depends on the partition $[n_I]$, or equivalently, it does not depend on the ordering of $(n_I)$.

The Levi subgroup $\mathbb{L}$ is a discrete label 
of the surface operator. 
This surface operator has $(M-1)$ continuous parameters $\alpha_{(M)}$ as we already saw.
We can turn on magnetic fluxes  $\mathfrak{m}^{I}$,  ($I=1,\ldots,M$), on the surface operator 
since the gauge group $\mathbb{L}$ on the surface operator has $(M-1)$ $\U(1)$ factors.
Correspondingly, there are parameters $\eta_{(I)}$,
giving rise to a phase factor $\exp (i \mathfrak{m}^I \eta_I)$ in the path integral.
Under supersymmetry, it is natural to combine $\vec\alpha$ and $\vec\eta$ to complex parameters 
$\vec{t} := \vec{\eta} + i \vec{\alpha}$.   
Half-BPS surface operators of $\cN=4$ theory has additional parameters $\vec\beta$ and $\vec\gamma$.

Let us introduce a few examples of surface operators.
When $\vec{\alpha}$ is the most generic, $\mathbb{L} = \U(1)^{N-1}$,
and the surface operator is called full. 
If $\vec{\alpha}\propto (\alpha,\alpha,\ldots,(1-N)\alpha )$, 
$\mathbb{L} = \SU(N-1) \times \U(1)$,
and the surface operator is called simple.
When $N=2$ the simple surface operator is the same as the full surface operator;
When  $N=3$  a nontrivial surface operator is either simple or full. 
If $\vec\alpha=0$, $\mathbb{L}=\SU(N)$, and this corresponds to the absence of the surface operator. 

Let  $\mathfrak{p}$ be a subalgebra of $\mathfrak{g}_{\bC}$ spanned by elements $x$ satisfying \begin{equation}
[\alpha,x] = i\lambda x, \qquad \lambda\ge 0.
\end{equation}
$\mathfrak{p}$ is called a parabolic subalgebra, and the corresponding subgroup $\cP\subset G_\bC$ is called a parabolic subgroup. 
As examples let us consider the case $N=4$ and $\bL=S[\U(2)\times\U(1)^2]$, corresponding to the partition $[2,1,1]$. 
When $\vec\alpha=(\alpha_{(1)},\alpha_{(1)},\alpha_{(2)},\alpha_{(3)})$ and the composition is $(2,1,1)$, 
$\cP$ is the subgroup generated by the elements of the form \begin{equation}
\left(\begin{array}{cc|c|c}
* & * & * & * \\
* & * & * & * \\
\hline 
0 & 0 & * & * \\
\hline 
0& 0 & 0 & *
\end{array}\right),
\end{equation}
and when  $\vec\alpha=(\alpha_{(1)},\alpha_{(2)},\alpha_{(2)},\alpha_{(3)})$ and the composition is $(1,2,1)$, 
$\cP$ is the subgroup generated by the elements of the form \begin{equation}
\left(\begin{array}{c|cc|c}
* & * & * & * \\
\hline 
0 & * & * & * \\
0 & * & * & * \\
\hline 
0& 0 & 0 & *
\end{array}\right) .
\end{equation}   These two parabolic subgroups are \emph{not} conjugate to each other inside $\SL(N)$, and the choice depends on the composition $(n_I)$ of $N$.
Note that when the Levi subgroup is $\bL=\U(1)^{N-1}$, there is a unique parabolic subgroup associated, consisting of all the semi-upper-triangular matrices. This parabolic subgroup is called the Borel subgroup and denoted by $\cB$.

The surface operator can also be described as a coupling
of the four-dimensional gauge theory with
a two-dimensional theory on the surface \cite{Gukov:2006jk}:
One can take a non-linear sigma model
whose target space is the partial flag manifold $G_\bC/\cP$. 
As Riemannian manifolds, $G_\bC/\cP$ is always the same with $G/\bL$,
but they can be different as complex manifolds, 
and therefore can give rise to different two-dimensional $\cN=(2,2)$ supersymmetric sigma models, see e.g.~\cite{Donagi:2007hi} for a recent discussion.
From this point of view, the parameters $\vec t$ are the complexified  K\"ahler parameters.
In contrast, a half-BPS surface operator of $\cN=4$ theory is  given by the coupling to the two-dimensional supersymmetric sigma model on $T^*(G_\bC/\cP)$. In this case the hyperK\"ahler structures for different $\cP$ with the same $\bL$ can be continuously interpolated by changing $(\vec\alpha,\vec\beta,\vec\gamma)$ \cite{Gukov:2006jk}.

\subsection{Instantons with a surface operator as orbifolded instantons}
We consider an instanton configuration $F=-{}\star F$ with the singularity \eqref{diverge} 
together with a choice of the composition $(n_1,n_2,\ldots,n_M)$ of $N$, see \eqref{ordered}.
The topological data are the instanton number $d$ and the magnetic fluxes $\mathfrak{m}^{I}$
of the $\U(1)$ factors of the Levi subgroup.  As always,  $\sum_I\mathfrak{m}^{I}=0$  because the overall $\U(1)$ is decoupled. 
To define the instanton number, we define a smooth gauge field $\bar A_\mu$ on $\bR^4$ by \begin{equation}
A_\mu dx^\mu=\bar A_\mu dx^\mu + f(r) \diag(\alpha_1,\ldots,\alpha_N) id\theta
\end{equation} where $f(r)$ is a smooth function that interpolates between $1$ at $w=0$ and $0$ at $w=\infty$. We define the instanton number $d$ of the configuration as the instanton number of $\bar A_\mu$. Then the integral of $\tr F\wedge F$ is given by \begin{equation}
\frac{1}{8\pi^2}\int_{\bC^2\setminus \{w=0\}} \tr F\wedge F  = d +\frac12 \sum_I \alpha_{(I)} \mathfrak{m}^{I}.
\end{equation}
The topological data are more conveniently parameterized  by $\vec{d}$ defined by \begin{equation}
d_M=d,\qquad d_{I+1}=d_I+\mathfrak{m}^I
\end{equation} where the index $I$ is taken modulo $M$.
The moduli space of such configurations is denoted by $\cM_{\bL,\vec\alpha,\vec d}$. 
When $d=0$, the moduli space describes excitations purely on the surface, given by the supersymmetric sigma model whose target space is $G_\bC/\cP$. This system was analyzed e.g.~in \cite{Braverman:2010ef,Dimofte:2010tz,Awata:2010bz,Bonelli:2011fq,Yoshida:2011au}. 

In mathematics literature, it is known \cite{MehtaSeshadri,Biswas} that the smooth part of this moduli space, as far as its complex structure is concerned, agrees with the smooth part of the moduli space of $\U(N)$ instantons on $\bC\times (\bC/\bZ_M)$ where $\bZ_M$ acts on the spacetime as \begin{equation}
(z,w) \to (z, \omega w)\label{equiv1}
\end{equation} and on the $N$-dimensional representation of $\U(N)$ as the multiplication by \begin{equation}
(
\underbrace{\omega,\ldots,\omega}_{\text{$n_1$ times}},
\underbrace{\omega^2,\ldots,\omega^2}_{\text{$n_2$ times}},
\ldots,
\underbrace{\omega^M,\ldots,\omega^M}_{\text{$n_M$ times}} )\label{equiv2}
\end{equation} where $\omega=e^{2\pi i/M}$.
As we will see shortly this equivalence allows us to describe the moduli space of instantons with a surface operator in terms of a quiver, obtained from a standard orbifold operation applied to the ADHM construction of the instantons on $\bC^2$. 

Before proceeding, we would like to sketch the derivation of the equivalence in two ways, one is mathematical and the other is string theoretical.
Mathematical derivation goes as follows.  
The singularities of the moduli space of differential-geometric objects can be resolved by translating them into algebro-geometric language. 
In this case, the corresponding objects in algebraic geometry is a rank-$N$ torsion-free sheaves on $\CP^1\times \CP^1$ with coordinates $(z,w)$, with framing given at $\{z=\infty\}\cup\{w=\infty\}$ and with parabolic structure of type $\cP$ given at $\{w=0\}$, 
whose precise definition can be found in \cite{FFNR,FR}.
Another standard mathematical trick, also described in \cite{FGK,FFNR,FR}, to study this moduli space is to use the one-to-one mapping between a parabolic sheaf on $\CP^1\times \CP^1$ of type $\cP$ and a $\bZ_M$-equivariant sheaf on $\CP^1\times \CP^1$, with precisely the conditions \eqref{equiv1}, \eqref{equiv2}.
This can be mapped back to a differential-geometric object as an anti-self-dual configuration on $\bC\times(\bC/\bZ_M)$, completing the derivation.

String theoretical derivation goes as follows. 
When we realize $\cN=4$ super Yang-Mills theory by compactifying the six-dimensional $\cN=(2,0)$ theory on $T^2$,
a surface operator of type $\bL$ in four dimensions can be obtained by placing a codimension-two defect of type $\bL$ in six dimensions along $T^2$. 
When the original theory is of type $A_{N-1}$ and can be realized as the low-energy theory of coincident $N$ M5-branes, it was shown in  \cite{Gaiotto:2009we} (also see Sec.~3 of \cite{Gaiotto:2009hg}) that the codimension-two defect of type $\bL$  corresponding to the partition $N=n_1+\cdots + n_M$ can be realized by considering a Type IIA configuration with $N$ D4-branes ending on $M$ D6-branes such that $n_I$  D4-branes end on the $I$-th D6-brane. 
Lifting the configuration to M-theory, we have $N$ M5-branes hitting the singularity of a multi Taub-NUT space.
As far as the complex structure is concerned, the situation is now described by the eleven-dimensional  spacetime
$\bR^2\times \bC\times (\bC^2/\bZ_M)\times \bR^3$ with coordinates $(x^0,x^1,z,w,w',x^7,x^8,x^9)$ where the orbifold is given by \begin{equation}
(w,w')\to (\omega w,\omega^{-1} w').
\end{equation}
 $N$ M5-branes are on $w'=x^7=x^8=x^9=0$, and the codimension-two defect is the orbifold singularity $w=0$. 
Compactifying this system along $x^0$ and $x^1$, we end up with $\cN=4$ Yang-Mills theory on the orbifold given in \eqref{equiv1}.

\subsection{Orbifolded instantons and the chain-saw quiver}
Let us recall how the moduli space of $\SU(N)$ instantons on $\bR^4$ with instanton number $d$ is given by the ADHM construction.
Let us introduce two vector spaces $V$ and $W$ with 
$\dim_{\mathbb C} V=d$ and $\dim_{\mathbb C} W= N$.
In the language of D-brane configuration the system we consider is
$d$ D0-branes bound to $N$ D4-branes.
As an effective theory on D4-branes we have $\U(N)$ gauge theory and 
$d$ D0-branes describe the gas of  $d$ point-like instantons. 
In this picture the ADHM construction is given by a dual description where
we consider an effective $0+1$-dimensional theory on D0 brane
\cite{Witten:1994tz, Witten:1995gx, Douglas:1995bn, Douglas:1996uz}.
We have $A, B \in \mathrm{Hom}~(V,V)$ from \lq\lq 0-0\rq\rq\ string.  
From \lq\lq 0-4\rq\rq\ and \lq\lq 4-0\rq\rq\ string
we have $P \in \mathrm{Hom}~(W,V)$ and $Q \in \mathrm{Hom}~(V,W)$.
The ADHM equation arises as the BPS condition for this  D-brane system 
and given by
\beqa
{\cal E}_{\mathbb C} &:=& [A, B] + PQ = 0~, \\
{\cal E}_{\mathbb R} &:=&[A, A^\dagger] + [B, B^\dagger] + P P^\dagger - Q^\dagger Q  =0~.
\eeqa
Then the moduli space is given by 
\beq
{\cal M}_{\mathrm{ADHM}} := \{ (A, B, P, Q) \vert~
{\cal E}_{\mathbb C}=0,~{\cal E}_{\mathbb R}=0 \}/ \U(d)~,
\eeq
where $g \in \U(d)$ acts on the ADHM data as follows;
\beq
(A, B, P, Q) \to (gAg^{-1}, gBg^{-1}, gP, Qg^{-1}).
\eeq
It is known that the resolution of singularity of the ADHM moduli space is given by the following affine algebro-geometric
quotient \cite{Nak, Nakajima:2003uh};
\beq
\widetilde{\cal M}_{\mathrm{ADHM}} := \{ (A, B, P, Q) \vert~
{\cal E}_{\mathbb C}=0,~\hbox{stability}  \}/\!/ \GL(d, {\mathbb C})~. \label{ADHMquotient}
\eeq
This is a smooth space of complex dimension $2Nd$.

To obtain the quiver describing the moduli space of instantons on $\bC\times (\bC/\bZ_M)$,
we perform the standard orbifolding procedure as in \cite{Kronheimer:1989zs,KronheimerNakajima,Douglas:1996sw}.
The vector spaces $V$ and $W$ are decomposed  according to the representation under $\bZ_M$ action,
\beq
W = {\bigoplus_{I = 1}^{M}} W_I, \qquad V = {\bigoplus_{I =1}^{M}} V_I.
\eeq
We also have \beq
\dim W_I = n_I, \qquad  \dim V_I = d_I. 
\eeq
In the language of branes, $W_I$ and $V_I$ are the Chan-Paton spaces of D4- and D0-branes on which the orbifold action acts as $\omega^I$. 
In particular, $d_I$ characterizes  the number of fractional instantons.

\begin{figure}[ht]
\begin{center}
\begin{pspicture}(-3,-1)(10,5) 
\rput(0,0){\circlenode{W1}{$W_{I -1}$}}
\rput(0,3){\circlenode{V1}{$V_{I -1}$}}
\rput(3,0){\circlenode{W2}{$~~W_{I ~}$}}
\rput(3,3){\circlenode{V2}{$~~V_{I ~}$}}
\rput(6,0){\circlenode{W3}{$W_{I +1}$}}
\rput(6,3){\circlenode{V3}{$V_{I +1}$}}
\rput(-2,0){\pnode{W0}}
\rput(-2,3){\pnode{V0}}
\rput(8,0){\pnode{W4}}
\rput(8,3){\pnode{V4}}
\nccircle[arrowsize=6pt]{->}{V1}{.7}
\nccircle[arrowsize=6pt]{->}{V2}{.7}
\nccircle[arrowsize=6pt]{->}{V3}{.7}

\ncline[arrowsize=6pt]{->}{V1}{V2}
\ncline[arrowsize=6pt]{->}{V2}{V3}

\ncline[arrowsize=6pt]{->}{W1}{V1}
\ncline[arrowsize=6pt]{->}{W2}{V2}
\ncline[arrowsize=6pt]{->}{W3}{V3}
\ncline[arrowsize=6pt]{->}{V1}{W2}
\ncline[arrowsize=6pt]{->}{V2}{W3}
\ncline[arrowsize=6pt, linestyle=dashed]{W0}{W1}
\ncline[arrowsize=6pt, linestyle=dashed]{V0}{V1}
\ncline[arrowsize=6pt, linestyle=dashed]{W3}{W4}
\ncline[arrowsize=6pt, linestyle=dashed]{V3}{V4}

\rput(0,4.7){$A_{I -1}$}
\rput(3,4.7){$A_{I }$}
\rput(6,4.7){$A_{I +1}$}
\rput(1.5,3.5){$B_{I -1}$}
\rput(4.5,3.5){$B_{I }$}
\rput(-0.5,1.5){$P_{I -1}$}
\rput(3.3,1.5){$P_{I }$}
\rput(6.5,1.5){$P_{I +1}$}
\rput(5,1.5){$Q_{I }$}
\rput(1,1.5){$Q_{I -1}$}
\end{pspicture}
\end{center}
\caption{A part of $\mathbb{Z}_M$ chain-saw quiver\label{chainsaw}}
\end{figure}
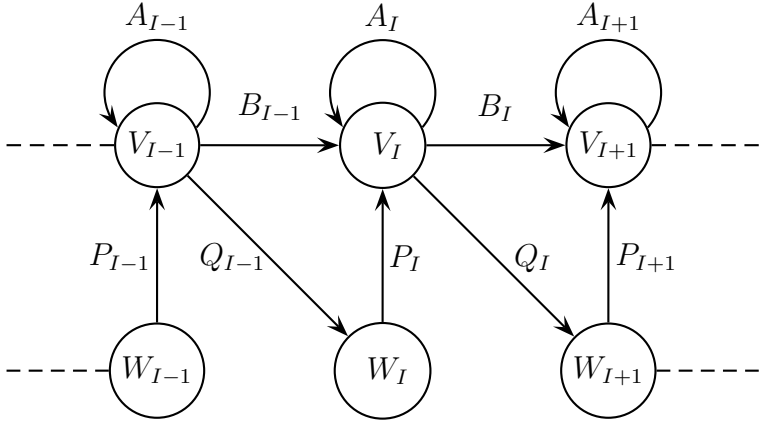

The isometry $(z, w) \to (e^{\ep} z, e^{\es} w)$ of ${\mathbb C}^2$
acts on the ADHM data as 
$(A, B, P, Q) \to ( e^{\ep}\cdot A, e^{\es} \cdot B, P, e^{\ep+\es}\cdot Q)$.
Therefore, under the $\mathbb{Z}_M$ orbifold action $(z,w)\to (z,\omega w)$,
the components which survive are 
$A_{I } \in \mathrm{Hom}~(V_I ,V_I )$,  $B_I  \in \mathrm{Hom}~(V_{I },V_{I +1})$,
$P_{I } \in \mathrm{Hom}~(W_I ,V_I )$ 
and $Q_{I } \in \mathrm{Hom}~(V_I ,W_{I +1})$, and
the equation of motion is
\beq
A_{I +1} B_{I } - B_{I } A_{I } + P_{I +1} Q_{I } = 0.
\eeq
Here in the following, the index $I$ is taken modulo $M$, therefore $V_{M+1}=V_1$, etc.
The resulting quiver is the chain-saw quiver  \cite{FR} shown in Fig.~\ref{chainsaw}. 
The resulting moduli space is a smooth complex manifold of dimension \begin{equation}
\sum_{I=1}^M  (n_I +n_{I +1}) d_I .
\end{equation}

\subsection{Instanton counting with a surface operator}
For  $\cN=2$ pure $\SU(N)$ gauge theory, it was shown in \cite{Nekrasov:2002qd} that the instanton contribution to the partition function can be expressed by the geometrical data specifying how the Cartan torus $\U(1)^2\times \U(1)^N$ of the symmetry $\SO(4)\times \U(N)$  acts on the instanton moduli space. We first recall how it was done for the case without the surface operator. Then the extension to the case with general surface operator is straightforward. 

\subsubsection{Before the orbifolding}
The action $(z, w) \to (e^{\ep} z, e^{\es} w)$ on ${\mathbb C}^2$
induces an action on the ADHM data given by \begin{equation}
(A, B, P, Q) \to ( e^{\ep}\cdot A, e^{\es} \cdot B, P, e^{\ep+\es}\cdot Q),
\end{equation}
and $g\in \U(N)$ acts as 
\begin{equation}
(A,B,P,Q) \to (A,B,Pg^{-1},gQ).
\end{equation}
Fixed points under these actions are isolated and  labeled by  $N$-tuples of Young diagrams $\vec\lambda=(\lambda^1,\ldots,\lambda^N)$; for a Young diagram $\lambda$ we denote the height of the $j$-th column by $\lambda_j$.

The ADHM data associated to an $N$-tuple $\vec\lambda$ is explicitly given as follows.
Let $w_1,\ldots,w_N$ a basis of $W$, such that $\diag(e^{a_1},\ldots,e^{a_N})\in  \U(N)$ acts by \begin{equation}
 w^I \to e^{a_I} w^I.
\end{equation}
Let us associate a basis $v_{(i,j)}^I$ of $V$ for 
each box $(i,j)\in\lambda^I$, i.e.~for a box at the $i$-th row, the $j$-th column of the Young diagram $\lambda^I$.
In particular, $\dim V$ is the total number of boxes in $\vec\lambda$.
For notational simplicity, we define $v_{(i,j)}^I=0$ if $(i,j)\not\in\lambda^I$.
The toric action on this basis is given by \begin{equation}
v_{(i,j)}^I \to e^{(1-i)\epsilon_1+(1-j)\epsilon_2} v_{(i,j)}^I.
\end{equation} 
We then have \begin{equation}
Av_{(i,j)}^I = v_{(i+1,j)}^I,\qquad
Bv_{(i,j)}^I = v_{(i,j+1)}^I,\qquad
Pw^I = v_{(1,1)}^I,\qquad
Q=0.
\end{equation}

The tangent space of the moduli space at the fixed point $\vec{\lambda}$ is given by the cohomology of the following complex \cite{Nak, Bruzzo:2002xf, Losev:2003py}:
\beq
\begin{array}{ccccc}
 & & \mathrm{Hom}~(V , V ) \otimes (T_1\oplus T_2) & & \\
 & & \oplus & & \\
\mathrm{Hom}~(V ,V ) &
~\overset{\sigma}{\longrightarrow}  & \mathrm{Hom}~(W ,V ) 
& \overset{\tau}{\longrightarrow}~&   \mathrm{Hom}~(V ,V ) \otimes T_1 \otimes T_2 \\ 
 & & \oplus & & \\
 & & \mathrm{Hom}~(V ,W ) \otimes T_1\otimes T_2 & & 
\end{array},\label{complex}
\eeq
where  $T_1$, $T_2$ are one dimensional modules on which $\U(1)^2$ 
act as the multiplication by $e^{\ep}$ and $e^{\es}$, respectively.
Hence the character of the tangent space at a fixed point $\vec\lambda$  under  $\U(1)^2\times\U(1)^{N}$ is given by
\begin{equation}
\chi_{\vec{\lambda}} =
 -  V^*  V (1- e^{\epsilon_1}) (1- e^{\epsilon_2}) 
 + W^*  V
+ V^*  W e^{\epsilon_1+\epsilon_2}  ~.\label{4Dcharacter}
\end{equation} Here, we abused the notation and identified the vector spaces and their characters, i.e.~\begin{equation}
V= \sum_I\sum_{(i,j)\in \lambda^I} e^{a_I +(1-i)\epsilon_1+(1-j)\epsilon_2}, \qquad
W=\sum_I  e^{a_I }
\end{equation} and the asterisk $*$ in the superscript reverses the sign of the exponents. 

In the expression \eqref{4Dcharacter},
all the terms with negative coefficient cancel out and 
there are $2Nd$ remaining terms, \begin{equation}
\chi_{\vec{\lambda}} = \sum_{i=1}^{2Nd} \exp(w_i(\vec\lambda)).
\end{equation}
Each weight $w_i$ is an integral linear combination of the equivariant parameters $\epsilon_{1,2}$ and $a_I $.
Then  Nekrasov's partition function of the pure $\SU(N)$  theory is  given by \begin{equation}
Z(\epsilon_{1,2},a_I,q)=\sum_{\vec\lambda} q^{|\vec \lambda|} z(\vec\lambda).
\end{equation} where
\beq
z(\vec{\lambda}) = \prod_{i=1}^{2Nd} \frac{1}{w_i(\vec\lambda)}~.
\eeq
In the physical terms, $q=\Lambda^{2N}$ is the dynamical scale, $a_I$ gives the vev of the adjoint scalar in the vector multiplet, and $\epsilon_{1,2}$ is the deformation parameter in Nekrasov's $\Omega$-background.

\subsubsection{After the orbifolding}
The fixed points after the orbifolding can be classified in a similar manner, and still labeled by $N$-tuples of Young diagrams, $\vec\lambda=(\lambda^{s,I })$ where $I=1,\ldots,M$,  $s=1,\ldots,n_I$.

We first redefine the equivariant parameters $\epsilon_{1,2}$ so that it acts on $(z, w)$ via \begin{equation}
(z,w)\to (e^{\ep} z, e^{\frac{1}{M}\es} w)
\end{equation} because the physical space corresponds to the wedge in the $w$-plane of the angle $2\pi/M$. 
We also rename the equivariant parameters $a_1,\ldots,a_N$ by $a_{s,I}$ where $I=1,\ldots,M$ and $s=1,\ldots,n_I$.

We  then introduce a basis $w_{s,I }$ of $W_I $ such that $\U(1)^2\times\U(1)^{N}$  acts by \begin{equation}
w^{s,I }\to e^{-\frac IM \epsilon_2 + a_{s,I }} w^{s,I }.
\end{equation} 
We also introduce a basis $v_{(i,j)}^{s,I +j}$ of $V_{I +j}$ for each $(i,j)\in\lambda^{s,I }$ such that $\U(1)^2\times \U(1)^N$ acts by \begin{equation}
v_{(i,j)}^{s,I +j} \to 
e^{-\frac IM \epsilon_2 + a_{s,I }}  e^{(1-i)\epsilon_1+\frac{1}M(1-j)\epsilon_2}
v_{(i,j)}^{s,I +j} 
\end{equation}

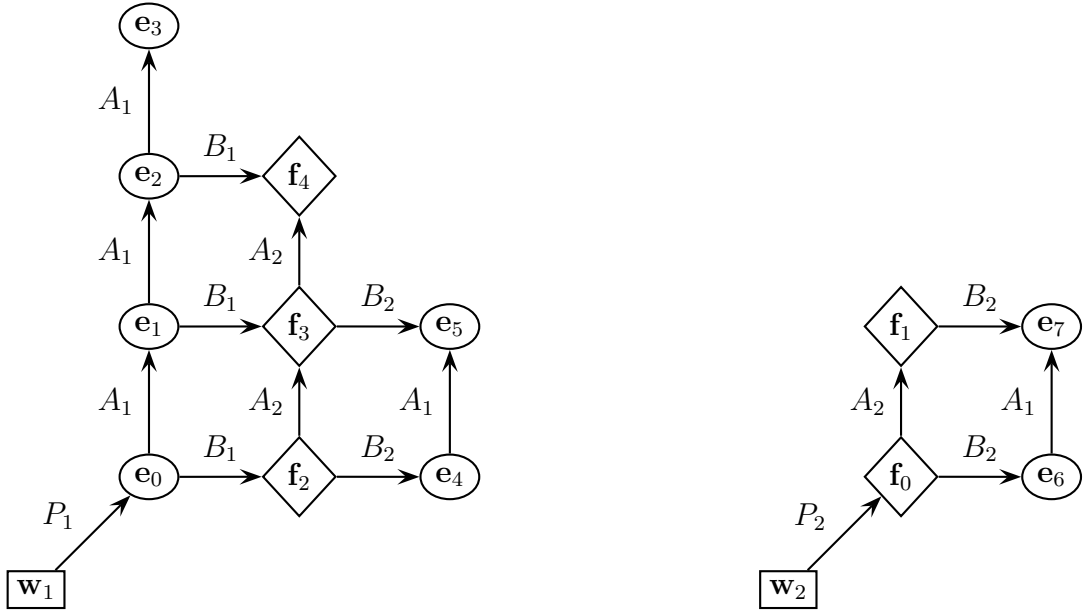
\begin{figure}[ht]

\begin{center}
\begin{pspicture}(-1.5,-1.5)(13,6) 
\rput(-1.5,-1.5){\rnode{V}{\psframebox{${\bf w}_1$}}}
\rput(0,0){\ovalnode{V11}{${\bf e}_0$}}
\rput(0,2){\ovalnode{V12}{${\bf e}_1$}}
\rput(0,4){\ovalnode{V13}{${\bf e}_2$}}
\rput(0,6){\ovalnode{V14}{${\bf e}_3$}}
\rput(2,0){\dianode{V21}{${\bf f}_2$}}
\rput(2,2){\dianode{V22}{${\bf f}_3$}}
\rput(2,4){\dianode{V23}{${\bf f}_4$}}
\rput(4,0){\ovalnode{V31}{${\bf e}_4$}}
\rput(4,2){\ovalnode{V32}{${\bf e}_5$}}
\rput(8.5,-1.5){\rnode{W}{\psframebox{${\bf w}_2$}}}
\rput(10,0){\dianode{W11}{${\bf f}_0$}}
\rput(10,2){\dianode{W12}{${\bf f}_1$}}
\rput(12,0){\ovalnode{W21}{${\bf e}_6$}}
\rput(12,2){\ovalnode{W22}{${\bf e}_7$}}
\ncline[arrowsize=6pt]{->}{V}{V11}
\naput[npos=.4]{$P_1$}
\ncline[arrowsize=6pt]{->}{V11}{V12}
\naput{$A_1$}
\ncline[arrowsize=6pt]{->}{V12}{V13}
\naput{$A_1$}
\ncline[arrowsize=6pt]{->}{V13}{V14}
\naput{$A_1$}
\ncline[arrowsize=6pt]{->}{V11}{V21}
\naput{$B_1$}
\ncline[arrowsize=6pt]{->}{V12}{V22}
\naput{$B_1$}
\ncline[arrowsize=6pt]{->}{V21}{V22}
\naput{$A_2$}
\ncline[arrowsize=6pt]{->}{V13}{V23}
\naput{$B_1$}
\ncline[arrowsize=6pt]{->}{V22}{V23}
\naput{$A_2$}
\ncline[arrowsize=6pt]{->}{V21}{V31}
\naput{$B_2$}
\ncline[arrowsize=6pt]{->}{V31}{V32}
\naput{$A_1$}
\ncline[arrowsize=6pt]{->}{V22}{V32}
\naput{$B_2$}
\ncline[arrowsize=6pt]{->}{W}{W11}
\naput[npos=.4]{$P_2$}
\ncline[arrowsize=6pt]{->}{W11}{W12}
\naput{$A_2$}
\ncline[arrowsize=6pt]{->}{W11}{W21}
\naput{$B_2$}
\ncline[arrowsize=6pt]{->}{W12}{W22}
\naput{$B_2$}
\ncline[arrowsize=6pt]{->}{W21}{W22}
\naput{$A_1$}
\end{pspicture}
\end{center}

\caption{A representation of the chain saw quiver for $(n_I)=(1,1)$,  $\lambda^1 = (4,3,2), \lambda^2 = (2,2)$
with $d_1 = 8$ and $d_2=5$. ${\bf e}_i$ forms the basis of $V_1$, ${\bf f}_i$ forms the basis of $V_2$, and
${\bf w}_{1,2}$ is the basis of $W_{1,2}$.  \label{examplefp}}
\end{figure}

Then the fixed configuration is given by 
\begin{equation}
A_I  v_{(i,j)}^{s,I } = v_{(i+1,j)}^{s,I },\quad
B_I  v_{(i,j)}^{s,I } = v_{(i,j+1)}^{s,I +1},\quad
P_I  w^{s,I } = v_{(1,1)}^{s,I },\quad
Q_I =0.
\end{equation}
Note  that the dimension of $V_{I }$ is then given by  \begin{equation}
d_{I }(\vec\lambda)=  \sum_{j=1}^\infty \sum_{s=1}^{n_{I+1-j}} \lambda^{s,I +1-j}_j,
\end{equation} i.e. the boxes in the $j$-th column of the Young diagram $\lambda^{s,I }$ contributes to the dimension of $V_{I +j-1}$.
To illustrate this point, an example of a fixed-point configuration for the $\bZ_2$ orbifold is shown in Fig.~\ref{examplefp}.

The character of the tangent space of the moduli space at $\vec\lambda$ is given by taking the $\bZ_M$ invariant part of the complex \eqref{complex}. The result is
\begin{equation}
\chi_{\vec\lambda}= \sum_{I=1}^M  \Bigl[ -(1-e^{\epsilon_1})  V_{I }^*  V_{I } 
+(1-e^{\epsilon_1})e^{\frac1M\epsilon_2} V_{I -1}^* V_{I } 
+ W_{I }^*  V_{I } 
+  e^{\epsilon_1+\frac1M\epsilon_2}  V_{I -1}^* W_{I }\Bigr]. \label{formula}
\end{equation} 
As before, we abused the notation and identified the vector space and its character, i.e.~
\begin{align}
V_I
& = \sum_{J=1}^{M}\sum_{s=1}^{n_{I-J+1}} e^{\lfloor \frac{I-J}M \rfloor \es - \frac{I}M\es+a_{s,I-J+1} } 
\!\!\!\sum_{(i, j M + J) \in \lam^{s,I-J+1}}\!\!\!
e^{(1-i) \ep - j \es} , \\
W_{I }&=\sum_{s=1}^{n_I } e^{-\frac{I}{M}\epsilon_2+a_{s,I }} ,
\end{align}
where $\lfloor \cdot \rfloor$ denotes the floor function. Since $1 \leq I,J \leq M$, it gives either $0$ for $I \geq J$ or $-1$ for $I < J$. Expansion of \eqref{formula} reproduces the conjecture made by Wyllard in \cite{Wyllard:2010vi}. 

After expansion, $\chi_{\vec\lambda}$ has the form
\begin{equation}
\chi_{\vec\lambda}=\sum_{i=1}^{\dim(\vec\lambda)} e^{w_i(\vec\lambda)}  
\end{equation} where $w_i$ are integral linear combinations of $a_{s,I }$ and $\epsilon_{1,2}$ and \begin{equation}
\dim(\vec\lambda)\equiv \sum (n_I +n_{I +1})d_I (\vec\lambda).
\end{equation} 
Then Nekrasov's partition function of the pure $\SU(N)$ gauge theory with a general surface operator  is  given by \begin{equation}
Z(\epsilon_{1,2},a_{s,I},q_I)=\sum_{\vec\lambda} z(\vec\lambda) \prod_{I } q_I ^{d_I (\vec\lambda)}.
\label{purepartitionfunction}
\end{equation} where
 \begin{equation}
z(\vec{\lambda}) = \prod_{i=1}^{\dim(\vec\lambda)} \frac{1}{w_i(\vec\lambda)}~.
\end{equation}

It is natural to assign mass dimension $1$ to the equivariant parameters $\epsilon_{1,2}$ and $a_{s,I }$. To make $Z$ dimensionless, we should then assign mass dimension $n_I +n_{I +1}$ to $q_I $. Then the mass dimension of \begin{equation}
\Lambda^{2N}\equiv q_1q_2\cdots q_M
\end{equation} is $2N$, and $\Lambda$ can be naturally identified with the dynamical scale of the gauge theory.

\section{Comparison with the W-algebra}\label{3}
\subsection{Structure of the W-algebra}\label{w}
The W-algebra we denote as $W(\hat\SU(N),[n_I])$ is obtained by the quantum Drinfeld-Sokolov reduction applied to the affine Lie algebra $\hat{\SU}(N)$ with an $\SU(2)$ embedding \begin{equation}
\rho:\SU(2)\to \SU(N),
\end{equation} where the fundamental $N$-dimensional representation of $\SU(N)$ is decomposed as  \begin{equation}
N=\underline{n_1}\oplus \underline{n_2}\oplus\cdots\oplus\underline{n_M}
\end{equation} under $\rho(\SU(2))$. Here $\underline{m}$ stands for an $m$-dimensional irreducible representation of $\SU(2)$.

The concrete procedure of the reduction is concisely explained in \cite{deBoer:1993iz}. 
For simplicity of exposition, we consider $W(\hat\U(N),[n_I])$ instead of $W(\hat\SU(N),[n_I])$.
The difference is a decoupled free boson.
The algebra is generated by fields \begin{equation}
U^{I}_{J,(s)}(z),   \quad 1\le I,J \le M, \quad s=\frac{|n_I-n_J|}2+1,\ldots ,\frac{|n_I+n_J|}2
\end{equation} where $s$ denotes the scaling dimension.
The Miura transformation realizes this algebra as a subalgebra of the current algebra \begin{equation}
\hat{\U}(m_1)\times
\hat{\U}(m_2)\times
\cdots
\hat{\U}(m_{M'})
\end{equation} where $[m_I]$ is the dual partition of $[n_I]$. 
The level of $\hat{\U}(m_I)$ is $k+N-m_I$, where $k$ is the level of the original $\hat\U(N)$ current algebra. It is also to be noted that $\hat\U(1)$ parts of $\hat{\U}(m_I)$ and $\hat{\U}(m_J)$ have non-zero off-diagonal components in the level matrix.

Let $\cI_n=\{I\,|\,n_I=n\}$. Then $\cJ^I_J(z)\equiv U^{I}_{J,(1)}(z)$ for $I,J\in \cI_n$ generate $\hat \U(|\cI_n|)$ current subalgebra of $W(\hat{\U}(N),[n_I])$, and fields with $s>1$ transform in the bifundamental representation under these current subalgebra. 

We consider the algebra in the R-sector, where every field has integer moding. Then the universal enveloping algebra is generated by operators $U^I_{J,(s),n}, n \in {\mathbb Z}$.
The Cartan subalgebra is generated by $N$ generators $U_{I,(s)}\equiv U^I_{I,(s),0}$  (no summation on $I$), and in particular 
$U_{I,(1)}$ satisfies \begin{equation}
[U_{I,(1)},U^J_{J',(s),n}]=(\delta_{IJ}-\delta_{IJ'})U^J_{J',(s),n}.
\end{equation}

We declare $U^I_{J,(s),n}$ with $I>J$, $n\ge 0$ and those with $I\le J$, $n>0$ to be the annihilation operators. 
The Verma module is then generated by the highest weight vector $\ket{u_{I,(s)}}$ where $u_{I,(s)}$ is the eigenvalue of $U_{I,(s)}$. 

Note that the structure of the Verma module depends on the composition $(n_I)$, not just the partition $[n_I]$.
To see this, let us assign a factor of $y_I^q$ for a state of charge $q$ under $U_{I,(1)}$,
and a factor of $z^n$ for a state at grade $n$, and call them the multiplicative charge of states or operators.
Then the multiplicative charge of $U^I_{J,(s),-n}$ is  $z^n y_J/y_I$. 
We define \begin{equation}
x_1=y_1/y_2, \ \ldots,\  x_{k-1}=y_{M-1}/y_M,\  x_M = z y_M/y_1.
\end{equation} 
Then the charge of every creation operator is a monomial with non-negative powers of $x_I$,
and the graded dimension of the Verma module is given by \begin{equation}
\tr \ z^{\Delta L_0} \prod_I y_I^{\Delta U_{I,(0)}}
=\prod_{i=1}^M \prod_{J=1}^\infty \left(1-\prod_{s=I}^{I+J-1} x_s \right)^{-\min(n_I,n_{I+J})}.\label{verma}
\end{equation} 
Here, $\Delta L_0$ and $\Delta U_{I,(1)}$ measure the difference of the respective conserved charges between the state concerned and the highest weight vector.
Thus the graded dimension clearly depends on the composition $(n_i)$.
Note that this is for the Verma module of $W(\hat\U(N),[n_I])$; the one for $W(\hat\SU(N),[n_I])$ is obtained by dropping a free boson, i.e.~dropping a factor of $\prod_n (1-(x_1\cdots x_M)^n)^{-1}$.

\subsection{Instantons and states in the Verma module}
Let us now compare the structure of the W-algebra with the instanton calculation.  
We consider first the five-dimensional maximally supersymmetric Yang-Mills with gauge group $\SU(N)$ on $\bR_t\times \bC_{z}\times \bC_{w}$, and introduce a codimension-two defect of type $\bL$ associated to a partition $[n_I]$, with monodromy $\vec\alpha$ at $w=0$, as defined in Sec.~\ref{2.1}. 
Its BPS sector can be reduced to a supersymmetric quantum mechanics on the instanton moduli spaces.  We perform Nekrasov's deformation, and then the Hilbert space of the BPS states is given by the equivariant cohomology \begin{equation}
\cH_\text{BPS}=\bigoplus_{d,\mathfrak{m}} H^*_{\U(1)^2\times \U(1)^{N-1}}(\cM_{\bL,\vec\alpha,\vec d}).
\end{equation}
This should also be the Hilbert space of the two-dimensional theory obtained by compactifying the six-dimensional $\cN=(2,0)$ theory on $\bC_{z}\times \bC_{w}$, further compactified on $S^1$. 
Then it is this two-dimensional theory that has the W-algebra symmetry $W(\hat\SU(N),[n_I])$.
Furthermore, $\cH_\text{BPS}$ is expected to become a Verma module  $\cV$ of the W-algebra, $\cH_\text{BPS}=\cV$.

We replace the instanton moduli space $\cM_{\bL,\vec\alpha,\vec d}$, which is highly singular,
 with its resolution of singularities, the moduli space of parabolic framed torsion free sheaves $\cM_{\cP,\vec d}$. 
 Here $\cP$ is the parabolic subgroup determined by $\vec\alpha$, or equivalently by the composition $(n_I)$.
Then we consider the Hilbert space \begin{equation}
\cH_\text{BPS}'=\bigoplus_{\vec d} H^*_{\U(1)^2\times \U(1)^N}(\cM_{\cP,\vec d}).
\end{equation}
Since the fixed points of the action of $\U(1)^2\times\U(1)^N$ on $\cM_{\cP,\vec d}$ are isolated, the equivariant localization theorem tells us that the equivariant cohomology has a basis labeled by the fixed points
\begin{equation}
\cH_\text{BPS}'=\bigoplus_{\vec\lambda} \bC \ket{\vec\lambda}.
\end{equation}

We expect that $\cH_\text{BPS}'$ is basically the same as the original Hilbert space $\cH_\text{BPS}$,
except for a factor $\cF$ from free bosons introduced by the resolution of singularities. Combining with the expectation $\cH_\text{BPS}=\cV_\cP$, we arrive at a general conjecture 
 \begin{equation}
\cH_\text{BPS}'= \cV_\cP\otimes \cF.
\end{equation} 

This statement has been checked for a few cases. When $N=1$, $\cH'_\text{BPS}$ consists solely of the Fock space of a free boson, and the action of the Heisenberg algebra was constructed geometrically in \cite{NakajimaHeisenberg,GrojnowskiHeisenberg,LehnHeisenberg}.
For $N>1$ with no surface operator, or equivalently when $\bL=\SU(N)$ and $\cP=\SL(N)$, $\cV_\cP$ is the Verma module of the $W_N$ algebra and $\cF$ is the Fock space of a free boson \cite{Alday:2009aq,Wyllard:2009hg,Alba:2010qc,Belavin:2011js}. When the surface operator is full, or equivalently when $\bL=\U(1)^{N-1}$ and $\cP$ is the Borel subgroup $\cB$, $\cV_\cP$ is the Verma module of the affine $\hat\SU(N)$ algebra and again $\cF$ is a Fock space \cite{Alday:2010vg,Kozcaz:2010yp}.  Moreover, in \cite{FGK,BFG}, the cohomology was studied directly before the resolution of singularities when $\cP=\cB$ and it was shown that they are the Verma module of $\hat\SL(N)$ without an extra  boson.

We would like to perform a few more checks of this proposal, for the surface operator of more general type. First, let us consider the generating function of the fixed points, \begin{equation}
\tr_{\cH'} \prod_I  x_I ^{d_I } =\sum_{\vec\lambda} \prod_I  x_I ^{d_I (\vec\lambda)},
\end{equation} which can be thought of as Nekrasov's partition function of $\cN=4$ super Yang-Mills theory,
because taking the trace means compactifying $\bR_t$ of the maximally supersymmetric five-dimensional Yang-Mills theory along a circle.

This generating function can be easily calculated by noticing that adding a row of length $j$ to $\lambda^{s,I }$ 
contributes a factor of $x_{I }x_{I +1}\cdots x_{I +j-1}$. Since there are $n_I $ choices of $s $, we find \begin{equation}
\tr_{\cH'}\prod_I  x_I ^{d_I }=
\prod_{I=1}^M \prod_{j=1}^\infty \left(1-\prod_{s=I}^{I+j-1} x_s \right)^{-n_I}.
\end{equation} Note that this is mostly the same as \eqref{verma} except that we have $n_I$ instead of $\min(n_I,n_{I+J-1})$ in the exponent.

In particular, when all of $n_I$ are equal, we see that the generating function gives the graded dimension of the Verma module of $W(\hat\U(N),\rho)$, or the product of the Verma module of $W(\hat\SU(N),\rho)$ and the Fock space of a free boson. This includes the cases referred to above,  $\bL=\SU(N)$ and $\bL=\U(1)^{N-1}$. It also covers the case when $n_I\le n_{I+1}$ for all $I$ and the quiver is not taken to be cyclic. This corresponds to the finite analogue dealt with in \cite{Braverman:2010ef}. 
However, it is clear that this is not the whole story. In the general case, one finds that in $\cH'$ there are  $1+\sum_{1\le I<J\le M} |n_I-n_J|$ bosons, presumably free, in addition to the Verma module $\cV_P$.

\subsection{Partition function and the Whittaker vector}
In order to study a more detailed structure of $\cH'_\text{BPS}$, let us consider the instanton partition function of the pure $\SU(N)$ gauge theory defined by \eqref{purepartitionfunction}. 
From the point of view of the five-dimensional maximally supersymmetric Yang-Mills, four-dimensional $\cN=2$ pure $\SU(N)$ theory is obtained by making $\bR_t$ into a segment with a common half-BPS boundary condition at both ends; recall Witten's construction using D4-branes suspended between NS5-branes \cite{Witten:1997sc}. 
The boundary condition determines a state $\ket W$ in $\cH_\text{BPS}$. Therefore, we expect the partition function to be given by the inner product, $Z=\vev{W|W}$, possibly up to a factor coming from the extra bosons in $\cF$.  The condition to be put on $\ket{W}$ can be guessed from the behavior of the Seiberg-Witten curve \cite{Gaiotto:2009ma} and is a certain coherent state, as detailed below. The state $\ket W$ is usually called the Whittaker vector in the representation theory.

This statement was checked  in \cite{Gaiotto:2009ma,Taki:2009zd} for the case without the surface operator, and proved for $\SU(2)$ in \cite{Fateev:2009aw,Marshakov:2010fx,Yanagida:2010kf,Yanagida:2010qm}. 
It was proved for the full surface operator of general $\SU(N)$ in \cite{Braverman:2004vv,Braverman:2004cr}. 
In these papers, it was always the case that for the pure gauge theory, the factor coming from the extra bosons was trivial.
Then it is natural to conjecture \cite{Wyllard:2010rp,Wyllard:2010vi} that for the pure gauge theory \begin{equation}
Z=\vev{W|W}
\label{Wnorm}\end{equation} where $\ket W$ is a coherent state in the Verma module, or the Whittaker vector. 
Note that for a general surface operator, both sides of \eqref{Wnorm}  depend on the choice of the parabolic subgroup,  or equivalently on the composition $(n_I)$. 

Let us determine the condition to be put on the coherent state. In general, a coherent state is defined by \begin{equation}
U^I_{J,(s),n} \ket W= u^I_{J,(s),n} \ket W
\end{equation} for annihilation operators $U^I_{J,(s),n}$. 
The ``eigenvalues'' $u^I_{J,(s),n}$ need to vanish whenever  $U^I_{J,(s),n}$ is given by a commutator of two other annihilation operators. 
One way to guarantee this condition is to allow nonzero  $u^I_{J,(s),n}$ only for the annihilation operators whose multiplicative charge as defined in Sec.~\ref{w} is $x_1$, $x_2$, \ldots, or $x_n$. 
For each $x_I$, we have annihilation operators $U^{I+1}_{I,(s),0}$ 
  where we define $U^{M+1}_{M,(s),0}\equiv U^1_{M,(s),1}$ for notational simplicity. 
Now, the expansion of $Z$ is in terms of $q_1$, $q_2$, \ldots, $q_M$ where $q_I$ has mass dimension $n_I + n_{I+1}$. 
As the mass dimension of the four-dimensional  theory is translated to the scaling dimension of the two-dimensional fields, which was given by $s$, the only sensible choice is to impose the conditions \begin{equation}
U^{I+1}_{I,(\frac{n_I+n_{I+1}}2),0} \ket W = c_I q_I^{1/2} \ket W
\end{equation}  and to let the remaining annihilation operators to annihilate $\ket W$. 
Here $c_I$ are constants depending on the precise normalization of the generators $U^I_{J,(s),n}$,
which we expect to be rational numbers when the generators are naturally defined via the quantum Drinfeld-Sokolov reduction. 
This condition is a natural generalization of the conditions proposed in \cite{Wyllard:2010rp,Wyllard:2010vi}, and also takes into account the dependence of the Verma module and the instanton partition function on the composition $(n_I)$. 

The equality $Z=\vev{W|W}$ was checked for a few cases by Wyllard \cite{Wyllard:2010rp,Wyllard:2010vi} using the explicit commutation relations of the corresponding W-algebras worked out for the partition of the form $[2,1,\ldots,1]$ in \cite{Bershadsky:1990bg,Romans:1990ta}.
The equality is also proved for the finite analogue in \cite{Braverman:2010ef} when $(n_I)$ is ordered so that $n_I \leq n_J$ when $I\leq J$. 

In the rest of the paper we will report the check of \eqref{Wnorm} for the partition 
\begin{equation}
[2^\mu1^\nu]=[\underbrace{2,\ldots,2}_\mu,\underbrace{1,\ldots,1}_\nu]
\end{equation}
 for various ordering. 
It is performed by first working out the quantum Drinfeld-Sokolov reduction and then implementing the resulting algebra in Mathematica.
The detail of the reduction is given in Appendix \ref{reduction} and the Mathematica file to calculate the instanton partition function and the Whittaker vector is available on the preprint webpage as an ancillary file so that any reader can check by herself/himself.

We start from $\hat\U(N)$ affine current algebra of level $k$.
The resulting W-algebra $W(\hat\SU(N),[2^\mu 1^\nu])$ has generators  \begin{equation}
j^a_b(z), \quad J^i_j(z), \quad
U^a_j(z), \quad \tilde U^i_b(z), \quad S^a_b(z)
\end{equation} where  $a,b\in \cI_2$ and $i,j\in \cI_1$. 
Here $\cI_n=\{I\, |\, n_I=n\}$, where $(n_I)$ is a composition of $N$ in terms of $\mu$ twos and $\nu$ ones.
These fields were respectively denoted by 
\begin{equation}
U^a_{b,(1)}(z),\quad 
U^i_{j,(1)}(z), \quad
U^a_{j,(3/2)}(z), \quad
U^i_{b,(3/2)}(z), \quad
U^a_{b,(2)}(z)
\end{equation} in Sec.~\ref{w}.
These are realized as a subalgebra of $\hat\U(\mu+\nu)_{k+\mu}\times \hat\U(\mu)_{k+\mu+\nu}$ current algebra, generated by fields $J^I_J(z)$ and $\check J^a_b(z)$
where $I,J=1,\ldots,\mu+\nu$ and $a,b\in \cI_2$.

Here we write down explicit forms of conditions put on the Whittaker vector, and the mapping of parameters between the W-algebra and the instanton counting  for a few explicit examples.
The highest weight of the Verma module is given by the eigenvalues of the Cartan of the W-generators, but the same information can be more conveniently encoded by the Cartan of $\hat\U(\mu+\nu)\times\hat\U(\mu)$, namely $h_I \equiv J^I_{I,0}$ and $\check h_a\equiv \check J^a_{a,0}$ (no summation on $I$ or $a$.) 
Then the conjecture is that \begin{equation}
Z(\epsilon_{1,2};a_i;q_I ) = \vev{W|W} \qquad \text{for  $\ket W\in \cV_{\cP,h_I,\check h_a}$}
\end{equation} 
under a suitable mapping between the parameters $\epsilon_{1,2}$, $a_i$ and $h_I,\check h_a$.

For the composition $(2,2,1,1)$,  the nonzero conditions we put on the Whittaker vector are 
\begin{equation}
\begin{array}{r@{}l@{\qquad}l@{}l}
S^2_{1,0}\ket W&=q_1^{1/2}\ket W , & \bra{W}S^{1}_{2,0} &= \bra{W}q_1^{1/2},\\
\tilde U^3_{2,0}\ket W&=q_2^{1/2}\ket W , & \bra{W}U^{2}_{3,0} &= \bra{W}q_2^{1/2},\\
j^4_{3,0}\ket W&=q_3^{1/2}\ket W , & \bra{W}j^{3}_{4,0} &= -\bra{W}q_3^{1/2},\\
U^0_{4,1}\ket W&=q_4^{1/2}\ket W , & \bra{W}\tilde U^{4}_{0,-1} &= -\bra{W}q_4^{1/2}.
\end{array}
\end{equation}
Then $Z$ equals $\vev{W|W}$ when  $k=-6-\epsilon_2/\epsilon_1$ and 
 \begin{equation}
(h_1,h_2,h_3,h_4,\check h_1,\check h_2) =
(\frac{a_{1,1}}{\epsilon_1},\frac{a_{1,2}}{\epsilon_1},\frac{a_{1,3}}{\epsilon_1},\frac{a_{1,4}}{\epsilon_1},
\frac{a_{2,1}+\epsilon_2}{\epsilon_1},\frac{a_{2,2}+\epsilon_2}{\epsilon_1})
+(1,2,3,4,5,6).
\end{equation}

For the composition $(2,1,2,1)$,  the nonzero conditions we put on the Whittaker vector are 
\begin{equation}
\begin{array}{r@{}l@{\qquad}l@{}l}
\tilde U^2_{1,0}\ket W&=q_1^{1/2}\ket W , & \bra{W}U^{1}_{2,0} &= \bra{W}q_1^{1/2},\\
U^3_{2,0}\ket W&=q_2^{1/2}\ket W , & \bra{W}\tilde U^{2}_{3,0} &= -\bra{W}q_2^{1/2},\\
\tilde U^4_{3,0}\ket W&=q_3^{1/2}\ket W , & \bra{W} U^{3}_{4,0} &= \bra{W}q_3^{1/2},\\
U^0_{4,1}\ket W&=q_4^{1/2}\ket W , & \bra{W}\tilde U^{4}_{0,-1} &= -\bra{W}q_4^{1/2}.
\end{array} 
\end{equation}
Then $Z$ equals $\vev{W|W}$ when  $k=-6-\epsilon_2/\epsilon_1$ and  \begin{equation}
(h_1,h_2,h_3,h_4,\check h_1,\check h_3) =
(\frac{a_{1,1}}{\epsilon_1},\frac{a_{1,2}}{\epsilon_1},\frac{a_{1,3}}{\epsilon_1},\frac{a_{1,4}}{\epsilon_1},
\frac{a_{2,1}+\epsilon_2}{\epsilon_1},\frac{a_{2,3}+\epsilon_2}{\epsilon_1})
+(1,2,3,4,5,6).
\end{equation}
Finally for the composition $(2,2,2)$, the nonzero conditions we put on the Whittaker vector are
\begin{equation}
\begin{array}{r@{}l@{\qquad}l@{}l}
S^2_{1,0}\ket W&=q_1^{1/2}\ket W , & \bra{W}S^{1}_{2,0} &= \bra{W}q_1^{1/2},\\
S^3_{2,0}\ket W&=q_2^{1/2}\ket W , & \bra{W}S^{2}_{3,0} &= \bra{W}q_2^{1/2},\\
S^0_{3,0}\ket W&=q_3^{1/2}\ket W , & \bra{W}S^{3}_{0,-1} &= \bra{W}q_3^{1/2}.
\end{array}
\end{equation}
Then $Z$ equals $\vev{W|W}$ when $k=-6-\epsilon_2/\epsilon_1$ and  \begin{multline}
(h_1,h_2,h_3,\check h_1,\check h_2,\check h_3) =\\
(\frac{a_{1,1}}{\epsilon_1},\frac{a_{1,2}}{\epsilon_1},\frac{a_{1,3}}{\epsilon_1},\frac{a_{2,1}+\epsilon_2}{\epsilon_1},
\frac{a_{2,2}+\epsilon_2}{\epsilon_1},\frac{a_{2,3}+\epsilon_2}{\epsilon_1})
+(1,2,3,4,5,6).
\end{multline} 

The general rule for the mapping is now clear. 
The salient points are that we always found the agreement, once the maps $k=-N-\epsilon_2/\epsilon_1$
and $h_i \sim a_i/\epsilon_1$ are used, and that the formula giving the level $k$ 
agrees with the previous papers \cite{Braverman:2004vv,Alday:2010vg,Kozcaz:2010yp,Wyllard:2010rp,Wyllard:2010vi}. 

\section{Conclusions}\label{4}
In this paper, we provided further checks of a recent conjecture that the instanton partition function of $\cN=2$ pure $\SU(N)$ gauge theory in the presence of a surface operator of type  $[n_I]$ is governed by the W-algebra $W(\hat\SU(N),[n_I])$. 
We first saw an explicit description of the instanton moduli space in the presence of a surface operator in terms of the representations of the chain-saw quiver and then determined the fixed point contributions. 
Then we studied the structure of the Verma module of the general W-algebra, and determined the conditions to be put on the Whittaker vector. 
We saw that the instanton partition function equals the norm of the Whittaker vector once an appropriate mapping between the parameters on the four-dimensional side and on the two-dimensional side was made. 
Along the way, we found that both the instanton moduli and the Verma module depend on the composition $(n_I)$, not just on the partition $[n_I]$.

Our explicit checks were only for surface operators of type $[2^\mu1^\nu]$. 
This was due to the technical problem that the quantum Drinfeld-Sokolov reduction for a partition $[n_I]$ involves normal-ordered products of $h$ currents where $h=\max(n_I)$. 
This makes the calculation very tedious when $h>2$.  
Then, one immediate generalization of this work would be to extend the checks to more general surface operators by implementing the quantum Drinfeld-Sokolov reduction itself on the computer. The authors would like to welcome the help of anybody interested. 
Another direction of study is to incorporate hypermultiplets to the analysis and to consider quiver gauge theories. 
This should correspond, on the W-algebra side, to the study of actions of primary fields on the Verma modules. 

In this paper we viewed the surface operator as coming from the codimension-two defect of the six-dimensional $\cN=(2,0)$ theory.
Therefore the presence of the surface operator changed the two-dimensional theory.
The surface operators of four-dimensional gauge theory can also be thought of as coming from the codimension-four defects of the six-dimensional theory.  
From this point of view, the introduction of a surface operator corresponds to an insertion of a degenerate field on  the Riemann surface, without changing the two-dimensional theory \cite{Alday:2009fs,Drukker:2010jp}. 
The latter viewpoint  was more suited to the comparison to the topological string approach in the case of the simple surface operator, as was exemplified in many papers \cite{Kozcaz:2010af,Dimofte:2010tz,Taki:2010bj,Awata:2010bz,Bonelli:2011fq}. 
Therefore it would be interesting to analyze general surface operators from this latter point of view. 
The relation of these two points of view should be related to the mapping between the Toda theory and the WZW theory \cite{Ribault:2005wp,Hikida:2007tq,Ribault:2008si,Teschner:2010je,Creutzig:2011qm}, and is also worth further investigation. In \cite{Bruzzo:2010fk} the instanton moduli space with a surface operator is described by a coupled system of the ADHM equation with two-dimensional vortex. It was  derived from the brane configuration realizing the codimension-four defects and called the enhanced ADHM system. A comparison of the enhanced ADHM quiver of \cite{Bruzzo:2010fk} and the chain-saw quiver may shed light on this issue. 

\section*{Acknowledgments}
The authors thank Hidetoshi Awata, Hiroyuki Fuji, Masahide Manabe, Satoshi Minabe, Soichi Okada, Yasuhiko Yamada and Shintaro Yanagida for discussions.
The authors cannot thank  Kentaro Nagao and Hiraku Nakajima too much for their explanations on various mathematical results. Without them this paper would have been impossible.
YT also thanks Daniel Green and Etsuko Itou for their help in Mathematica calculations. 
YT thanks the hospitality of Nagoya University during the GCOE spring school ``Gauge Theory, Gravity, and String Theory'' and during the ``Mini-workshop on Moduli of Instantons.''
The work of HK is supported in part by Grant-in-Aid for Scientific Research
[\# 22224001] from MEXT, Japan.
The work of YT is supported in part by NSF grant PHY-0969448  and by the Marvin L. Goldberger membership through the Institute for Advanced Study.
He was also supported in part by World Premier International Research Center Initiative (WPI Initiative),  MEXT, Japan through the Institute for the Physics and Mathematics of the Universe, the University of Tokyo.

\appendix 

\section{Details of the Drinfeld-Sokolov reduction}\label{reduction}

We perform the quantum Drinfeld-Sokolov reduction \cite{Bershadsky:1989mf,Feigin:1990pn,deBoer:1993iz} for the partition \begin{equation}
N=\underbrace{2+\cdots+2}_\mu + \underbrace{1+\cdots+1}_\nu,
\end{equation}
 following the procedure described in \cite{deBoer:1993iz}.  
For an extensive list of references, see \cite{Bouwknegt:1992wg,Bouwknegt:1995ag}.
We consider $\U(N)$ instead of $\SU(N)$. 

First, an index of the adjoint of $\U(N)$ is denoted by $^A_B$ where $A,B=1,\ldots,N$. This corresponds to a single index $a$ in the paper \cite{deBoer:1993iz}. The structure constant is then \begin{equation}
f^{{}^A_B\ {}^C_D}{}_{^E_F}=-\delta^C_B\delta^A_E\delta^F_D +\delta^A_D\delta^F_B\delta^C_E.
\end{equation} It is too cumbersome to keep this notation, so we use the convention \begin{equation}
T^A_B\equiv T^{^A_B} \equiv T_{^B_A}
\end{equation} then the left-hand side of the equation above is just $f^{ACF}_{BDE}$.

We start from the $\U(N)$ currents $J^A_B$; our unhatted currents are the hatted currents of \cite{deBoer:1993iz}. 
The $\SU(2)$ embedding for the partition $[2^\mu1^\nu]$  is given by \begin{equation}
t_0=\diag(\underbrace{\frac12,\ldots,\frac12}_\mu,\underbrace{0,\ldots,0}_\nu,\underbrace{-\frac12,\ldots,-\frac12}_\mu)
\end{equation} and the grading is done by \begin{equation}
\delta=\diag(\underbrace{1,\ldots,1}_\mu,\underbrace{1,\ldots,1}_\nu,\underbrace{0,\ldots,0}_\mu ).
\end{equation} 
Accordingly, we decompose $J^A_B$ as follows : \begin{equation}
J^A_B=\left(\begin{array}{c|c|c}
\tilde\jmath^a_b & U^a_j & V^a_b \\
\hline
X^i_b & J^i_j & Y^i_b \\
\hline 
\pmb{1} & 0& \check \jmath^a_b
\end{array}\right) \label{decomposition}
\end{equation} where $a,b=1,\ldots,\mu$ and $i,j=1,\ldots,\nu$.
Also we introduce $I,J=1,\ldots,\mu+\nu$ and let \begin{equation}
J^I_J=\left(\begin{array}{c|c}
\tilde\jmath^a_b & U^a_j \\
\hline
X^i_b & J^i_j  
\end{array}\right), \qquad
Y^I_b = \left(\begin{array}{c} V^a_b \\ \hline Y^i_b 
\end{array}\right). \label{miura1}
\end{equation}
The $c$ ghosts are \begin{equation}
c^I_b =\left( \begin{array}{c} c^a_b \\ \hline c^i_b \end{array}\right).
\end{equation}

The Miura transformation realizes the W-algebra as a subalgebra of $\hat\U(\mu+\nu)_{k+\mu}\times \hat\U(\mu)_{k+\mu+\nu}$ 
generated by $J^I_J$ and $\check \jmath^a_b$.
Note that there are also  off-diagonal components in the level matrix: \begin{equation}
J^I_J(z) \check\jmath^a_b(w) \sim \delta^I_J \delta^a_b \frac{1}{(z-w)^2}.\label{off}
\end{equation}

The generators of the W-algebras are \begin{equation}
j^a_b = \tilde\jmath^a_b+\check\jmath^a_b, \quad
J^i_j, \quad
U^a_i \label{miura2}
\end{equation} and \begin{equation}
S^a_b \equiv V^a_b {}^{(1)} ,\quad
\tilde U^i_b \equiv Y^i_b {}^{(1)} 
\end{equation} where $B^{(1)}$ is a solution to the tic-tac-toe \begin{equation}
D_0(B^{(1)}) = D_1 B.
\end{equation} 
At this stage it is already clear that the W-algebra has a current subalgebra $\U(\mu)_{2k+N}\times\U(\nu)_{k+\mu}$.

To determine $S^a_b $ and $\tilde U^i_b $ explicitly, 
we need the BRST differentials. The differential $D_0$ is given by \begin{equation}
D_0(X^i_b)=-c^i_b,\quad
D_0(\tilde\jmath^a_b)=-c^a_b,\quad
D_0(\check\jmath^a_b)=+c^a_b
\end{equation} and the differential $D_1$ is given by \begin{equation}
D_1(Y^I_b)=-\check\jmath^a_b c^I_a + J^I_J c^J_b + (k+N) \partial c^I_b 
\end{equation} which can be further decomposed to \begin{align}
D_1(V^a_b)&=-\check\jmath^c_b c^a_c +\tilde\jmath^a_c c^c_b + U^a_jc^j_b +(k+N) \partial c^a_b, \\
D_1(Y^i_b)&=-\check\jmath^a_b c^i_a +X^i_a c^a_b + J^i_jc^j_b +(k+N) \partial c^i_b. 
\end{align}

Therefore we find \begin{align}
S^a_b &= -U^a_j X^j_b + \check\jmath^c_b \tilde\jmath^a_c - (k+\mu+\nu)\partial \frac12(\tilde\jmath^a_b-\check\jmath^a_b), \label{miura3}\\
\tilde U^i_b &= \check\jmath^a_b X^i_a - J^i_j X^j_b - (k+\mu+\nu)\partial X^i_b.\label{miura4}.
\end{align}
In the expressions above, the normal ordering is implicit in the product of the currents, i.e. $(JJ)_m=\sum_i : J_i J_{m-i}: $  where  \begin{equation}
: J_n J_m: = \left\{ \begin{array}{ll}
J_n J_m & \text{if $n\le -1$}, \\
J_m J_n & \text{if $n\ge 0$}.
\end{array} \right.
\end{equation} Note that this is the OPE normal ordering and not the naive normal ordering.

Summarizing, the W-algebra has affine $\U(\mu)_{2k+N}\times\U(\nu)_{k+\mu}$ currents $
j^a_b 
, 
J^i_j,
$ the ``bifundamentals''  $
U^a_i, 
 \tilde U^i_a
$ and spin-two fields $
S^a_b,
$ written in terms of affine $\hat\U(\mu+\nu)_{k+\mu}\times \hat\U(\mu)_{k+\mu+\nu}$ currents $J^I_J$ and $\check \jmath^a_b$  via the relations \eqref{miura1},\eqref{off},\eqref{miura2},\eqref{miura3} and \eqref{miura4}.

For $m=n=1$, $N=3$, the W-algebra is the Bershadsky-Polyakov algebra $W^{(2)}_3$. 
Our formula reproduces (3.16) of de Boer-Tjin \cite{deBoer:1993iz}, by assigning their names (3.14) to the components given in \eqref{decomposition}, except for \begin{equation}
S=W^8_1 + \frac{\hat J^4 \hat J^4}{36}-\frac{\partial \hat J^4}2.
\end{equation} This discrepancy is just a redefinition of the basis in the algebra.

\bibliographystyle{ytphys}
\small\baselineskip=.8\baselineskip
\bibliography{paper}{}

\providecommand{\href}[2]{#2}\begingroup\raggedright\begin{thebibliography}{10}

\bibitem{Moore:1997dj}
G.~W. Moore, N.~Nekrasov, and S.~Shatashvili, ``{Integrating over Higgs
  Branches},'' \href{http://dx.doi.org/10.1007/PL00005525}{{\em Commun. Math.
  Phys.} {\bfseries 209} (2000) 97--121},
\href{http://arxiv.org/abs/hep-th/9712241}{{\ttfamily arXiv:hep-th/9712241}}.

\bibitem{Nekrasov:2002qd}
N.~A. Nekrasov, ``{Seiberg-Witten Prepotential from Instanton Counting},'' {\em
  Adv. Theor. Math. Phys.} {\bfseries 7} (2004) 831--864,
\href{http://arxiv.org/abs/hep-th/0206161}{{\ttfamily arXiv:hep-th/0206161}}.

\bibitem{Losev:2003py}
A.~S. Losev, A.~Marshakov, and N.~A. Nekrasov, ``{Small Instantons, Little
  Strings and Free Fermions},'' in {\em {From Fields to Strings:
  Circumnavigating Theoretical Physics, Ian Kogan Memorial Collection}},
  M.~Shifman, A.~Vainshtein, and J.~Wheater, eds., pp.~581--621.
\newblock World Scientific, 2005.
\newblock
\href{http://arxiv.org/abs/hep-th/0302191}{{\ttfamily arXiv:hep-th/0302191}}.
\newblock

\bibitem{Nekrasov:2003rj}
N.~Nekrasov and A.~Okounkov, ``{Seiberg-Witten Theory and Random Partitions},''
  in {\em {The Unity of Mathematics}}, P.~Etingof, V.~Retakh, and I.~M. Singer,
  eds., vol.~244 of {\em {Progress in Mathematics}}.
\newblock {Birkh{\"auser}}, 2006.
\newblock
\href{http://arxiv.org/abs/hep-th/0306238}{{\ttfamily arXiv:hep-th/0306238}}.
\newblock

\bibitem{Alday:2009aq}
L.~F. Alday, D.~Gaiotto, and Y.~Tachikawa, ``{Liouville Correlation Functions
  from Four-Dimensional Gauge Theories},''
  \href{http://dx.doi.org/10.1007/s11005-010-0369-5}{{\em Lett. Math. Phys.}
  {\bfseries 91} (2010) 167--197},
\href{http://arxiv.org/abs/0906.3219}{{\ttfamily arXiv:0906.3219 [hep-th]}}.

\bibitem{Wyllard:2009hg}
N.~Wyllard, ``{$A_{N-1}$ Conformal Toda Field Theory Correlation Functions from
  Conformal ${\mathcal{N}}\!=2$ $SU(N)$ Quiver Gauge Theories},''
  \href{http://dx.doi.org/10.1088/1126-6708/2009/11/002}{{\em JHEP} {\bfseries
  11} (2009) 002},
\href{http://arxiv.org/abs/0907.2189}{{\ttfamily arXiv:0907.2189 [hep-th]}}.

\bibitem{Mironov:2009by}
A.~Mironov and A.~Morozov, ``{On AGT Relation in the Case of U(3)},''
  \href{http://dx.doi.org/10.1016/j.nuclphysb.2009.09.011}{{\em Nucl. Phys.}
  {\bfseries B825} (2010) 1--37},
\href{http://arxiv.org/abs/0908.2569}{{\ttfamily arXiv:0908.2569 [hep-th]}}.

\bibitem{Gaiotto:2009ma}
D.~Gaiotto, ``{Asymptotically Free ${\mathcal{N}}\!=2$ Theories and Irregular
  Conformal Blocks},''
\href{http://arxiv.org/abs/0908.0307}{{\ttfamily arXiv:0908.0307 [hep-th]}}.

\bibitem{Marshakov:2009gn}
A.~Marshakov, A.~Mironov, and A.~Morozov, ``{On Non-Conformal Limit of the AGT
  Relations},'' \href{http://dx.doi.org/10.1016/j.physletb.2009.10.077}{{\em
  Phys. Lett.} {\bfseries B682} (2009) 125--129},
\href{http://arxiv.org/abs/0909.2052}{{\ttfamily arXiv:0909.2052 [hep-th]}}.

\bibitem{Taki:2009zd}
M.~Taki, ``{On AGT Conjecture for Pure Super Yang-Mills and W- Algebra},''
\href{http://arxiv.org/abs/0912.4789}{{\ttfamily arXiv:0912.4789 [hep-th]}}.

\bibitem{Bonelli:2009zp}
G.~Bonelli and A.~Tanzini, ``{Hitchin Systems, ${\mathcal{N}}\!=2$ Gauge
  Theories and W-Gravity},''
\href{http://arxiv.org/abs/0909.4031}{{\ttfamily arXiv:0909.4031 [hep-th]}}.

\bibitem{Kanno:2009ga}
S.~Kanno, Y.~Matsuo, S.~Shiba, and Y.~Tachikawa, ``{${\mathcal{N}}\!=2$ Gauge
  Theories and Degenerate Fields of Toda Theory},''
\href{http://arxiv.org/abs/0911.4787}{{\ttfamily arXiv:0911.4787 [hep-th]}}.

\bibitem{Drukker:2010jp}
N.~Drukker, D.~Gaiotto, and J.~Gomis, ``{The Virtue of Defects in 4D Gauge
  Theories and 2D CFTs},''
\href{http://arxiv.org/abs/1003.1112}{{\ttfamily arXiv:1003.1112 [hep-th]}}.

\bibitem{Passerini:2010pr}
F.~Passerini, ``{Gauge Theory Wilson Loops and Conformal Toda Field Theory},''
  \href{http://dx.doi.org/10.1007/JHEP03(2010)125}{{\em JHEP} {\bfseries 03}
  (2010) 125},
\href{http://arxiv.org/abs/1003.1151}{{\ttfamily arXiv:1003.1151 [hep-th]}}.

\bibitem{Kanno:2010kj}
S.~Kanno, Y.~Matsuo, and S.~Shiba, ``{Analysis of Correlation Functions in Toda
  Theory and AGT-W Relation for $SU(3)$ Quiver},''
  \href{http://dx.doi.org/10.1103/PhysRevD.82.066009}{{\em Phys. Rev.}
  {\bfseries D82} (2010) 066009},
\href{http://arxiv.org/abs/1007.0601}{{\ttfamily arXiv:1007.0601 [hep-th]}}.

\bibitem{Drukker:2010vg}
N.~Drukker and F.~Passerini, ``{(De)Tails of Toda CFT},''
\href{http://arxiv.org/abs/1012.1352}{{\ttfamily arXiv:1012.1352 [hep-th]}}.

\bibitem{Hollands:2010xa}
L.~Hollands, C.~A. Keller, and J.~Song, ``{From So/Sp Instantons to W-Algebra
  Blocks},'' \href{http://dx.doi.org/10.1007/JHEP03(2011)053}{{\em JHEP}
  {\bfseries 03} (2011) 053},
\href{http://arxiv.org/abs/1012.4468}{{\ttfamily arXiv:1012.4468 [hep-th]}}.

\bibitem{Braverman:2004vv}
A.~Braverman, ``{Instanton Counting via Affine Lie Algebras I: Equivariant
  J-Functions of (Affine) Flag Manifolds and Whittaker Vectors},'' in {\em
  {Workshop on algebraic structures and moduli spaces: CRM Workshop}},
  J.~Hurturbise and E.~Markman, eds.
\newblock AMS, July, 2003.
\newblock
\href{http://arxiv.org/abs/math/0401409}{{\ttfamily arXiv:math/0401409}}.
\newblock

\bibitem{Braverman:2004cr}
A.~Braverman and P.~Etingof, ``{Instanton Counting via Affine Lie Algebras. II:
  from Whittaker Vectors to the Seiberg-Witten Prepotential},'' in {\em Studies
  in Lie Theory: dedicated to A.~Joseph on his 60th birthday}, J.~Bernstein,
  V.~Hinich, and A.~Melnikov, eds.
\newblock {Birkh\"auser}, 2006.
\newblock
\href{http://arxiv.org/abs/math/0409441}{{\ttfamily arXiv:math/0409441}}.
\newblock

\bibitem{Alday:2010vg}
L.~F. Alday and Y.~Tachikawa, ``{Affine $SL(2)$ Conformal Blocks from 4D Gauge
  Theories},'' \href{http://dx.doi.org/10.1007/s11005-010-0422-4}{{\em Lett.
  Math. Phys.} {\bfseries 94} (2010) 87--114},
\href{http://arxiv.org/abs/1005.4469}{{\ttfamily arXiv:1005.4469 [hep-th]}}.

\bibitem{Kozcaz:2010yp}
C.~Koz\c{c}az, S.~Pasquetti, F.~Passerini, and N.~Wyllard, ``{Affine $SL(N)$
  Conformal Blocks from ${\mathcal{N}}\!=2$ $SU(N)$ Gauge Theories},''
  \href{http://dx.doi.org/10.1007/JHEP01(2011)045}{{\em JHEP} {\bfseries 01}
  (2011) 045},
\href{http://arxiv.org/abs/1008.1412}{{\ttfamily arXiv:1008.1412 [hep-th]}}.

\bibitem{Braverman:2010ef}
A.~Braverman, B.~Feigin, M.~Finkelberg, and L.~Rybnikov, ``{A Finite Analog of
  the AGT Relation I: Finite W-Algebras and Quasimaps' Spaces},''
\href{http://arxiv.org/abs/1008.3655}{{\ttfamily arXiv:1008.3655 [math.AG]}}.

\bibitem{Bershadsky:1989mf}
M.~Bershadsky and H.~Ooguri, ``{Hidden $SL(N)$ Symmetry in Conformal Field
  Theories},''
\href{http://dx.doi.org/10.1007/BF02124331}{{\em Commun. Math. Phys.}
  {\bfseries 126} (1989) 49}.

\bibitem{Feigin:1990pn}
B.~Feigin and E.~Frenkel, ``{Quantization of the Drinfeld-Sokolov Reduction},''
\href{http://dx.doi.org/10.1016/0370-2693(90)91310-8}{{\em Phys. Lett.}
  {\bfseries B246} (1990) 75--81}.

\bibitem{deBoer:1993iz}
J.~de~Boer and T.~Tjin, ``{The Relation Between Quantum W-Algebras and Lie
  Algebras},'' \href{http://dx.doi.org/10.1007/BF02103279}{{\em Commun. Math.
  Phys.} {\bfseries 160} (1994) 317--332},
\href{http://arxiv.org/abs/hep-th/9302006}{{\ttfamily arXiv:hep-th/9302006}}.

\bibitem{Wyllard:2010rp}
N.~Wyllard, ``{W-Algebras and Surface Operators in ${\mathcal{N}}\!=2$ Gauge
  Theories},'' \href{http://dx.doi.org/10.1088/1751-8113/44/15/155401}{{\em J.
  Phys.} {\bfseries A44} (2011) 155401},
\href{http://arxiv.org/abs/1011.0289}{{\ttfamily arXiv:1011.0289 [hep-th]}}.

\bibitem{Wyllard:2010vi}
N.~Wyllard, ``{Instanton Partition Functions in ${\mathcal{N}}\!=2$ $SU(N)$
  Gauge Theories with a General Surface Operator, and Their W-Algebra Duals},''
  \href{http://dx.doi.org/10.1007/JHEP02(2011)114}{{\em JHEP} {\bfseries 02}
  (2011) 114},
\href{http://arxiv.org/abs/1012.1355}{{\ttfamily arXiv:1012.1355 [hep-th]}}.

\bibitem{Tachikawa:2011dz}
Y.~Tachikawa, ``{On W-Algebras and the Symmetries of Defects of 6D N=(2,0)
  Theory},'' \href{http://dx.doi.org/10.1007/JHEP03(2011)043}{{\em JHEP}
  {\bfseries 03} (2011) 043},
\href{http://arxiv.org/abs/1102.0076}{{\ttfamily arXiv:1102.0076 [hep-th]}}.

\bibitem{FFNR}
B.~Feigin, M.~Finkelberg, A.~Negut, and L.~Rybnikov, ``{Yangians and cohomology
  rings of Laumon spaces},''
  \href{http://dx.doi.org/10.1007/s00029-011-0059-x}{{\em Selecta Mathematica}
  {\bfseries 17} (2008) 1--35},
  \href{http://arxiv.org/abs/0812.4656}{{\ttfamily arXiv:0812.4656 [math.AG]}}.

\bibitem{FR}
M.~Finkelberg and L.~Rybnikov, ``{Quantization of Drinfeld Zastava},''
  \href{http://arxiv.org/abs/1009.0676}{{\ttfamily arXiv:1009.0676 [math.AG]}}.

\bibitem{Gukov:2006jk}
S.~Gukov and E.~Witten, ``{Gauge Theory, Ramification, and the Geometric
  Langlands Program},''
  \href{http://projecteuclid.org/euclid.cdm/1223654541}{{\em Current
  Development in Mathematics} (2006) 35--180},
\href{http://arxiv.org/abs/hep-th/0612073}{{\ttfamily arXiv:hep-th/0612073}}.

\bibitem{Gukov:2007ck}
S.~Gukov, ``{Surface Operators and Knot Homologies},''
  \href{http://dx.doi.org/10.1002/prop.200610385}{{\em Fortschr. Phys.}
  {\bfseries 55} (2007) 473--490},
\href{http://arxiv.org/abs/0706.2369}{{\ttfamily arXiv:0706.2369 [hep-th]}}.

\bibitem{Gaiotto:2009fs}
D.~Gaiotto, ``{Surface Operators in ${\mathcal{N}}\!=2$ 4D Gauge Theories},''
\href{http://arxiv.org/abs/0911.1316}{{\ttfamily arXiv:0911.1316 [hep-th]}}.

\bibitem{Tan:2009he}
M.-C. Tan, ``{Notes on the `Ramified' Seiberg-Witten Equations and
  Invariants},''
\href{http://arxiv.org/abs/0912.1891}{{\ttfamily arXiv:0912.1891 [hep-th]}}.

\bibitem{Tan:2009qq}
M.-C. Tan, ``{Integration over the $u$-plane in Donaldson Theory with Surface
  Operators},''
\href{http://arxiv.org/abs/0912.4261}{{\ttfamily arXiv:0912.4261 [hep-th]}}.

\bibitem{Tan:2010dk}
M.-C. Tan, ``{Supersymmetric Surface Operators, Four-Manifold Theory and
  Invariants in Various Dimensions},''
\href{http://arxiv.org/abs/1006.3313}{{\ttfamily arXiv:1006.3313 [hep-th]}}.

\bibitem{Bruzzo:2010fk}
U.~Bruzzo, W.-Y. Chuang, D.-E. Diaconescu, M.~Jardim, G.~Pan, and Y.~Zhang,
  ``{D-Branes, Surface Operators, and ADHM Quiver Representations},''
\href{http://arxiv.org/abs/1012.1826}{{\ttfamily arXiv:1012.1826 [hep-th]}}.

\bibitem{Donagi:2007hi}
R.~Donagi and E.~Sharpe, ``{GLSM's for Partial Flag Manifolds},''
  \href{http://dx.doi.org/10.1016/j.geomphys.2008.07.010}{{\em J. Geom. Phys.}
  {\bfseries 58} (2008) 1662--1692},
\href{http://arxiv.org/abs/0704.1761}{{\ttfamily arXiv:0704.1761 [hep-th]}}.

\bibitem{Dimofte:2010tz}
T.~Dimofte, S.~Gukov, and L.~Hollands, ``{Vortex Counting and Lagrangian
  3-Manifolds},''
\href{http://arxiv.org/abs/1006.0977}{{\ttfamily arXiv:1006.0977 [hep-th]}}.

\bibitem{Awata:2010bz}
H.~Awata, H.~Fuji, H.~Kanno, M.~Manabe, and Y.~Yamada, ``{Localization with a
  Surface Operator, Irregular Conformal Blocks and Open Topological String},''
\href{http://arxiv.org/abs/1008.0574}{{\ttfamily arXiv:1008.0574 [hep-th]}}.

\bibitem{Bonelli:2011fq}
G.~Bonelli, A.~Tanzini, and J.~Zhao, ``{Vertices, Vortices \& Interacting
  Surface Operators},''
\href{http://arxiv.org/abs/1102.0184}{{\ttfamily arXiv:1102.0184 [hep-th]}}.

\bibitem{Yoshida:2011au}
Y.~Yoshida, ``{Localization of Vortex Partition Functions in
  $\mathcal{N}=(2,2)$ Super Yang-Mills Theory},''
\href{http://arxiv.org/abs/1101.0872}{{\ttfamily arXiv:1101.0872 [hep-th]}}.

\bibitem{MehtaSeshadri}
V.~B. Mehta and C.~S. Seshadri, ``Moduli of vector bundles on curves with
  parabolic structures,'' \href{http://dx.doi.org/10.1007/BF01420526}{{\em
  Mathematische Annalen} {\bfseries 248} (1980) 205--239}.

\bibitem{Biswas}
I.~Biswas, ``Parabolic bundles as orbifold bundles,''
  \href{http://dx.doi.org/10.1215/S0012-7094-97-08812-8}{{\em Duke Math.
  Journal} {\bfseries 88} (1997) 305--325}.

\bibitem{FGK}
M.~Finkelberg, D.~Gaitsgory, and A.~Kuznetsov, ``{Uhlenbeck spaces for
  $\mathbb{A}^2$ and affine Lie algebra $\hat{sl}_n$},''
  \href{http://dx.doi.org/10.2977/prims/1145476045}{{\em Pub. of the Research
  Inst. for Math. Sci.} {\bfseries 39} (2003) 721},
  \href{http://arxiv.org/abs/math.AG/0202208}{{\ttfamily
  arXiv:math.AG/0202208}}.

\bibitem{Gaiotto:2009we}
D.~Gaiotto, ``{${\mathcal{N}}\!=2$ Dualities},''
\href{http://arxiv.org/abs/0904.2715}{{\ttfamily arXiv:0904.2715 [hep-th]}}.

\bibitem{Gaiotto:2009hg}
D.~Gaiotto, G.~W. Moore, and A.~Neitzke, ``{Wall-Crossing, Hitchin Systems, and
  the WKB Approximation},''
\href{http://arxiv.org/abs/0907.3987}{{\ttfamily arXiv:0907.3987 [hep-th]}}.

\bibitem{Witten:1994tz}
E.~Witten, ``{Sigma Models and the ADHM Construction of Instantons},''
  \href{http://dx.doi.org/10.1016/0393-0440(94)00047-8}{{\em J. Geom. Phys.}
  {\bfseries 15} (1995) 215--226},
\href{http://arxiv.org/abs/hep-th/9410052}{{\ttfamily arXiv:hep-th/9410052}}.

\bibitem{Witten:1995gx}
E.~Witten, ``{Small Instantons in String Theory},''
  \href{http://dx.doi.org/10.1016/0550-3213(95)00625-7}{{\em Nucl. Phys.}
  {\bfseries B460} (1996) 541--559},
\href{http://arxiv.org/abs/hep-th/9511030}{{\ttfamily arXiv:hep-th/9511030}}.

\bibitem{Douglas:1995bn}
M.~R. Douglas, ``{Branes Within Branes},''
\href{http://arxiv.org/abs/hep-th/9512077}{{\ttfamily arXiv:hep-th/9512077}}.

\bibitem{Douglas:1996uz}
M.~R. Douglas, ``{Gauge Fields and D-Branes},''
  \href{http://dx.doi.org/10.1016/S0393-0440(97)00024-7}{{\em J. Geom. Phys.}
  {\bfseries 28} (1998) 255--262},
\href{http://arxiv.org/abs/hep-th/9604198}{{\ttfamily arXiv:hep-th/9604198}}.

\bibitem{Nak}
H.~Nakajima, {\em Lecture on Hilbert Scheme of Points on Surfaces}, vol.~18 of
  {\em University Lecture Series}.
\newblock AMS, 1999.

\bibitem{Nakajima:2003uh}
H.~Nakajima and K.~Yoshioka, ``{Lectures on Instanton Counting},'' in {\em
  {Workshop on algebraic structures and moduli spaces: CRM Workshop}},
  J.~Hurturbise and E.~Markman, eds.
\newblock AMS, July, 2003.
\newblock
\href{http://arxiv.org/abs/math/0311058}{{\ttfamily arXiv:math/0311058}}.
\newblock

\bibitem{Kronheimer:1989zs}
P.~B. Kronheimer, ``{The Construction of ALE Spaces as hyperk\"ahler
  Quotients},''
\href{http://projecteuclid.org/euclid.jdg/1214443066}{{\em J. Diff. Geom.}
  {\bfseries 29} (1989) 665--683}.

\bibitem{KronheimerNakajima}
P.~B. Kronheimer and H.~Nakajima, ``{Yang-Mills instantons on ALE gravitational
  instantons},'' \href{http://dx.doi.org/10.1007/BF01444534}{{\em Mathematische
  Annalen} {\bfseries 288} (1990) 263--307}.

\bibitem{Douglas:1996sw}
M.~R. Douglas and G.~W. Moore, ``{D-Branes, Quivers, and ALE Instantons},''
\href{http://arxiv.org/abs/hep-th/9603167}{{\ttfamily arXiv:hep-th/9603167}}.

\bibitem{Bruzzo:2002xf}
U.~Bruzzo, F.~Fucito, J.~F. Morales, and A.~Tanzini, ``{Multi-Instanton
  Calculus and Equivariant Cohomology},'' {\em JHEP} {\bfseries 05} (2003) 054,
\href{http://arxiv.org/abs/hep-th/0211108}{{\ttfamily arXiv:hep-th/0211108}}.

\bibitem{NakajimaHeisenberg}
H.~Nakajima, ``{Heisenberg Algebra and Hilbert Schemes of Points on Projective
  Surfaces},'' \href{http://www.jstor.org/stable/2951818}{{\em Annals of
  Mathematics} {\bfseries 145} (1997) 379--388}.

\bibitem{GrojnowskiHeisenberg}
I.~Grojnowski, ``{Instantons and affine algebras I: The Hilbert scheme and
  vertex operators},''
  \href{http://www.mathjournals.org/mrl/1996-003-002/1996-003-002-012.html}{{\em
  Math. Research Letters} {\bfseries 3} (1996) 275--291}.

\bibitem{LehnHeisenberg}
M.~Lehn, ``Chern classes of tautological sheaves on hilbert schemes of points
  on surfaces,'' \href{http://dx.doi.org/10.1007/s002220050307}{{\em Invent.
  Math.} {\bfseries 136} (1997) 157--207}.

\bibitem{Alba:2010qc}
V.~A. Alba, V.~A. Fateev, A.~V. Litvinov, and G.~M. Tarnopolsky, ``{On
  Combinatorial Expansion of the Conformal Blocks Arising from AGT
  Conjecture},''
\href{http://arxiv.org/abs/1012.1312}{{\ttfamily arXiv:1012.1312 [hep-th]}}.

\bibitem{Belavin:2011js}
A.~Belavin and V.~Belavin, ``{AGT Conjecture and Integrable Structure of
  Conformal Field Theory for $c=1$},''
\href{http://arxiv.org/abs/1102.0343}{{\ttfamily arXiv:1102.0343 [hep-th]}}.

\bibitem{BFG}
A.~Braverman, M.~Finkelberg, and D.~Gaitsgory, ``{Uhlenbeck spaces via affine
  Lie algebras},'' in {\em {The Unity of Mathematics: in Honor of the Ninetieth
  Birthday of I.~M.~Gelfand}}, P.~Etingof, V.~Retakh, and I.~M. Singer, eds.
\newblock {Birkh\"auser}, 2006.
\newblock \href{http://arxiv.org/abs/math.AG/0301176}{{\ttfamily
  arXiv:math.AG/0301176}}.

\bibitem{Witten:1997sc}
E.~Witten, ``{Solutions of Four-Dimensional Field Theories via M-Theory},''
  \href{http://dx.doi.org/10.1016/S0550-3213(97)00416-1}{{\em Nucl. Phys.}
  {\bfseries B500} (1997) 3--42},
\href{http://arxiv.org/abs/hep-th/9703166}{{\ttfamily arXiv:hep-th/9703166}}.

\bibitem{Fateev:2009aw}
V.~A. Fateev and A.~V. Litvinov, ``{On AGT Conjecture},''
  \href{http://dx.doi.org/10.1007/JHEP02(2010)014}{{\em JHEP} {\bfseries 02}
  (2010) 014},
\href{http://arxiv.org/abs/0912.0504}{{\ttfamily arXiv:0912.0504 [hep-th]}}.

\bibitem{Marshakov:2010fx}
A.~Marshakov, A.~Mironov, and A.~Morozov, ``{On AGT Relations with Surface
  Operator Insertion and Stationary Limit of Beta-Ensembles},''
  \href{http://dx.doi.org/10.1016/j.geomphys.2011.01.012}{{\em J. Geom. Phys.}
  {\bfseries 61} (2011) 1203--1222},
\href{http://arxiv.org/abs/1011.4491}{{\ttfamily arXiv:1011.4491 [hep-th]}}.

\bibitem{Yanagida:2010kf}
S.~Yanagida, ``{Whittaker Vectors of the Virasoro Algebra in Terms of Jack
  Symmetric Polynomial},''
\href{http://arxiv.org/abs/1003.1049}{{\ttfamily arXiv:1003.1049 [math.QA]}}.

\bibitem{Yanagida:2010qm}
S.~Yanagida, ``{Norms of Logarithmic Primaries of Virasoro Algebra},''
\href{http://arxiv.org/abs/1010.0528}{{\ttfamily arXiv:1010.0528 [math.QA]}}.

\bibitem{Bershadsky:1990bg}
M.~Bershadsky, ``{Conformal Field Theories via Hamiltonian Reduction},''
\href{http://dx.doi.org/10.1007/BF02102729}{{\em Commun. Math. Phys.}
  {\bfseries 139} (1991) 71--82}.

\bibitem{Romans:1990ta}
L.~Romans, ``{Quasisuperconformal algebras in two-dimensions and Hamiltonian
  reduction},'' \href{http://dx.doi.org/10.1016/0550-3213(91)90481-C}{{\em
  Nucl.Phys.} {\bfseries B357} (1991) 549--564}.

\bibitem{Alday:2009fs}
L.~F. Alday, D.~Gaiotto, S.~Gukov, Y.~Tachikawa, and H.~Verlinde, ``{Loop and
  Surface Operators in ${\mathcal{N}}\!=2$ Gauge Theory and Liouville Modular
  Geometry},'' \href{http://dx.doi.org/10.1007/JHEP01(2010)113}{{\em JHEP}
  {\bfseries 01} (2010) 113},
\href{http://arxiv.org/abs/0909.0945}{{\ttfamily arXiv:0909.0945 [hep-th]}}.

\bibitem{Kozcaz:2010af}
C.~Koz\c{c}az, S.~Pasquetti, and N.~Wyllard, ``{A \& B Model Approaches to
  Surface Operators and Toda Theories},''
  \href{http://dx.doi.org/10.1007/JHEP08 (2010)042}{{\em JHEP} {\bfseries 08}
  (2010) 042},
\href{http://arxiv.org/abs/1004.2025}{{\ttfamily arXiv:1004.2025 [hep-th]}}.

\bibitem{Taki:2010bj}
M.~Taki, ``{Surface Operator, Bubbling Calabi-Yau and AGT Relation},''
\href{http://arxiv.org/abs/1007.2524}{{\ttfamily arXiv:1007.2524 [hep-th]}}.

\bibitem{Ribault:2005wp}
S.~Ribault and J.~Teschner, ``{$H_3^+$ WZNW Correlators from Liouville
  Theory},'' \href{http://dx.doi.org/10.1088/1126-6708/2005/06/014}{{\em JHEP}
  {\bfseries 06} (2005) 014},
\href{http://arxiv.org/abs/hep-th/0502048}{{\ttfamily arXiv:hep-th/0502048}}.

\bibitem{Hikida:2007tq}
Y.~Hikida and V.~Schomerus, ``{$H^+_3$ WZNW Model from Liouville Field
  Theory},'' \href{http://dx.doi.org/10.1088/1126-6708/2007/10/064}{{\em JHEP}
  {\bfseries 10} (2007) 064},
\href{http://arxiv.org/abs/0706.1030}{{\ttfamily arXiv:0706.1030 [hep-th]}}.

\bibitem{Ribault:2008si}
S.~Ribault, ``{On $SL(3)$ Knizhnik-Zamolodchikov Equations and $W_3$
  Null-Vector Equations},''
  \href{http://dx.doi.org/10.1088/1126-6708/2009/10/002}{{\em JHEP} {\bfseries
  10} (2009) 002},
\href{http://arxiv.org/abs/0811.4587}{{\ttfamily arXiv:0811.4587 [hep-th]}}.

\bibitem{Teschner:2010je}
J.~Teschner, ``{Quantization of the Hitchin Moduli Spaces, Liouville Theory,
  and the Geometric Langlands Correspondence I},''
\href{http://arxiv.org/abs/1005.2846}{{\ttfamily arXiv:1005.2846 [hep-th]}}.

\bibitem{Creutzig:2011qm}
T.~Creutzig, Y.~Hikida, and P.~B. R{\o}nne, ``{Supergroup -- Extended Super
  Liouville Correspondence},''
\href{http://arxiv.org/abs/1103.5753}{{\ttfamily arXiv:1103.5753 [hep-th]}}.

\bibitem{Bouwknegt:1992wg}
P.~Bouwknegt and K.~Schoutens, ``{W Symmetry in Conformal Field Theory},''
  \href{http://dx.doi.org/10.1016/0370-1573(93)90111-P}{{\em Phys. Rept.}
  {\bfseries 223} (1993) 183--276},
\href{http://arxiv.org/abs/hep-th/9210010}{{\ttfamily arXiv:hep-th/9210010}}.

\bibitem{Bouwknegt:1995ag}
P.~Bouwknegt and K.~Schoutens, eds., {\em {W-Symmetry}}, vol.~22 of {\em
  Advanced Series in Mathemathical Physics}.
\newblock World Scientific,
1995.
\newblock

\end{thebibliography}\endgroup

\end{document}